\documentclass[10pt]{article} 
\usepackage[accepted]{tmlr}

\usepackage{amsmath,amsfonts,bm}

\def\eqref#1{equation~\ref{#1}}

\def\1{\bm{1}}

\def\rp{{\textnormal{p}}}

\def\rvc{{\mathbf{c}}}

\def\rvo{{\mathbf{o}}}

\def\rvx{{\mathbf{x}}}
\def\rvy{{\mathbf{y}}}
\def\rvz{{\mathbf{z}}}

\def\rmC{{\mathbf{C}}}

\def\rmP{{\mathbf{P}}}

\DeclareMathAlphabet{\mathsfit}{\encodingdefault}{\sfdefault}{m}{sl}
\SetMathAlphabet{\mathsfit}{bold}{\encodingdefault}{\sfdefault}{bx}{n}

\DeclareMathOperator*{\argmax}{arg\,max}

\DeclareMathOperator*{\length}{len}

\usepackage{amssymb}
\usepackage{hyperref}
\usepackage{url}
\usepackage{array}
\usepackage{balance}
\usepackage{latexsym}
\usepackage[T1]{fontenc}
\usepackage{microtype}
\usepackage{wrapfig}
\usepackage{graphicx}
\usepackage[table]{xcolor}
\usepackage{xfp}
\usepackage{amsmath}
\usepackage{booktabs}
\usepackage{multicol}
\usepackage{multirow}
\usepackage{capt-of}
\usepackage{fontawesome5}
\usepackage{newfloat}
\usepackage{scalerel}
\usepackage[most]{tcolorbox}
\usepackage{caption}
\usepackage{subcaption}
\usepackage{enumitem}
\usepackage{tablefootnote}
\usepackage{float}
\usepackage{pifont}
\usepackage{titlesec}
\usepackage{enumitem}
\usepackage{fancyvrb}
\usepackage{adjustbox}
\usepackage{varwidth}
\usepackage{placeins}
\usepackage{transparent}
\usepackage[noabbrev,capitalize,nameinlink]{cleveref}
\usepackage[varqu,varl]{inconsolata}
\titlespacing*{\section}{0pt}{1.2ex plus .2ex minus .4ex}{1.2ex plus .1ex}
\titlespacing*{\subsection}{0pt}{1.2ex plus .2ex minus .4ex}{1.2ex plus .1ex}
\definecolor{highlightblue}{HTML}{4C8BF5}
\definecolor{promptblue}{HTML}{292BA3}
\definecolor{weakorange}{HTML}{FF9D00}
\definecolor{strongorange}{HTML}{FF5100}
\definecolor{easygreen}{HTML}{40BF53}
\definecolor{hardgreen}{HTML}{14870C}
\definecolor{deepseekpurple}{HTML}{0D28F3}
\definecolor{olmopink}{HTML}{F0529c}

\newcommand{\workname}{{\texttt{Aletheia}}}
\definecolor{bonblue}{HTML}{4C8BF5}
\definecolor{extgreen}{HTML}{34A853}
\newcommand{\scval}[1]{\cellcolor{teal!\fpeval{round(max(0, (#1-25)*1.1))}}{#1}}
\newcommand{\rcval}[1]{\cellcolor{orange!\fpeval{round(max(0, (#1-5)*1.2))}}{#1}}
\newcommand{\miscval}[1]{\cellcolor{highlightblue!\fpeval{round(max(0, (#1-15)*0.8))}}{#1}}
\newcommand{\bonval}[1]{\cellcolor{bonblue!\fpeval{round(max(0, (#1-16)*9))}}{#1}}
\newcommand{\extval}[1]{\cellcolor{extgreen!\fpeval{round(max(0, (#1-25)*1.1))}}{#1}}

\newcommand{\ci}[1]{\textsubscript{\,#1}}
\newcommand{\scvalc}[2]{\scval{#1}\ci{#2}}
\newcommand{\rcvalc}[2]{\rcval{#1}\ci{#2}}
\newcommand{\bonvalc}[2]{\bonval{#1}\ci{#2}}
\newcommand{\extvalc}[2]{\extval{#1}\ci{#2}}

\newcolumntype{H}{w{c}{1.1cm}}
\newcolumntype{S}{w{c}{1cm}}
\newcolumntype{D}{w{c}{0.8cm}}
\newcolumntype{A}{w{c}{0.8cm}}
\newcolumntype{M}{w{c}{0.8cm}}
\newcommand{\train}{\texttt{\workname-Train}}
\newcommand{\heldout}{\texttt{\workname-Heldout}}
\newcommand{\dpo}{\texttt{\workname-DPO}}
\newcommand{\rqone}{\texttt{\workname-Strong}}
\newcommand{\rqtwo}{\texttt{\workname-Hard}}
\newcommand{\rqfour}{\texttt{\workname-Adv}}
\newcommand{\mixed}{\texttt{\workname-Mixed}}
\newcommand{\tablecross}{{\faTimes}}
\newcommand{\tabletick}{{\faCheckCircle}}
\newcommand{\weak}{\textcolor{weakorange}{\textbf{\texttt{Weak}}}}
\newcommand{\sstrong}{\textcolor{strongorange}{\textbf{\texttt{Strong}}}}
\newcommand{\easy}{\textcolor{easygreen}{\textbf{\texttt{Easy}}}}
\newcommand{\hard}{\textcolor{hardgreen}{\textbf{\texttt{Hard}}}}
\newcommand{\ukp}{\raisebox{-0.25\height}{\includegraphics[height=1em]{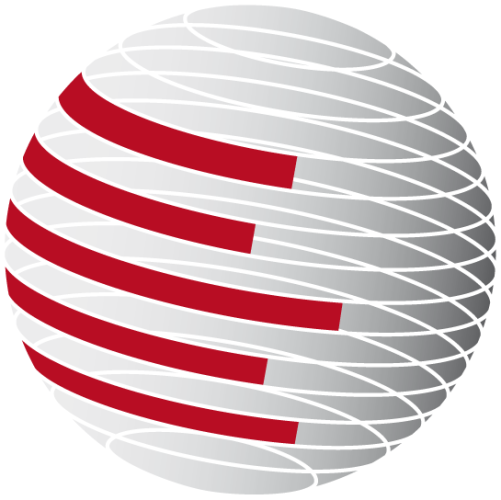}}}
\newcommand{\insait}{\raisebox{-0.25\height}{\includegraphics[height=1em]{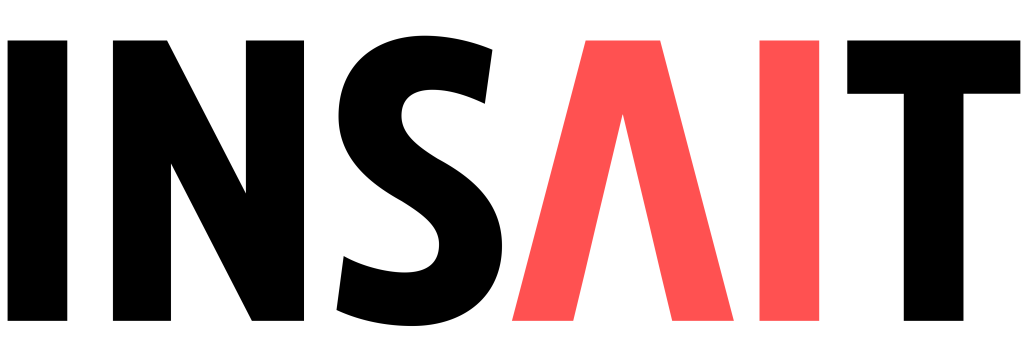}}}
\newcommand{\thinking}{\raisebox{0\height}{\includegraphics[height=0.9em]{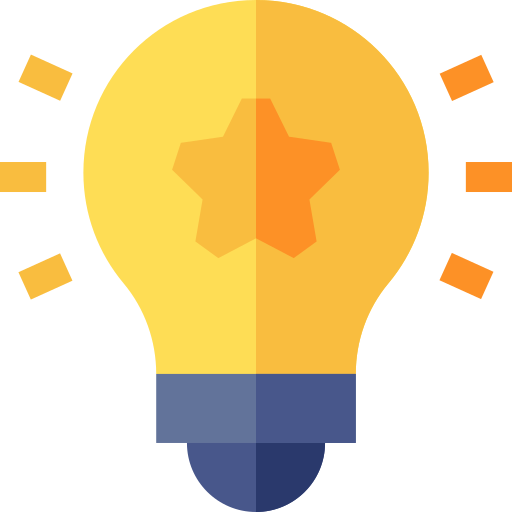}}}
\newcommand{\online}{\raisebox{0\height}{\includegraphics[height=0.9em]{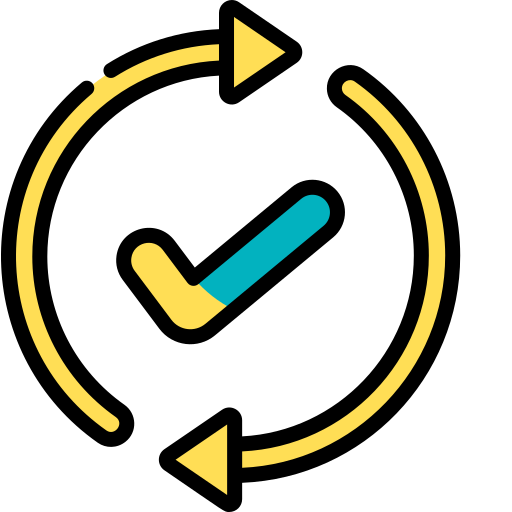}}}
\newcommand{\semionline}{\raisebox{0\height}{\includegraphics[height=0.9em]{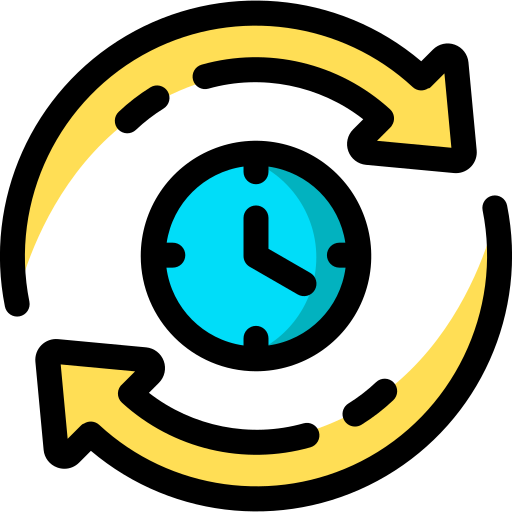}}}
\newcommand{\negatives}{\raisebox{0\height}{\includegraphics[height=0.9em]{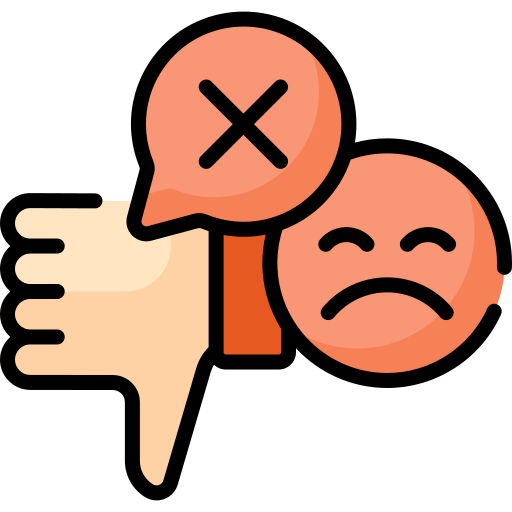}}}
\newcommand{\nothinking}{\raisebox{0\height}{\transparent{0.2}\includegraphics[height=0.9em]{Figures/thinking.png}}}
\newcommand{\nonegatives}{\raisebox{0\height}{\transparent{0.2}\includegraphics[height=0.9em]{Figures/negatives.png}}}
\newcommand{\noonline}{\raisebox{0\height}{\transparent{0.2}\includegraphics[height=0.9em]{Figures/online.png}}}

\newcommand{\ktau}{\texttt{K$\tau$}}

\newtcolorbox[auto counter]{finding}[1][]{
  colback=highlightblue!3!white,
  colframe=highlightblue!55!black,
  fonttitle=\scshape\small\bfseries,
  title=Summary Of Findings~\thetcbcounter,
  enhanced,
  boxrule=0.5pt,
  left=1mm,right=1mm,top=1mm,bottom=1mm,
  #1
}
\newtcolorbox[auto counter]{finding_bon}[1][]{
  colback=teal!3!white,
  colframe=teal!70!black,
  fonttitle=\tt\small,
  title={\textbf{Summary Of Findings For Best-of-N~\thetcbcounter}},
  enhanced,
  boxrule=0.5pt,
  left=1mm,right=1mm,top=1mm,bottom=1mm,
  #1
}
\newtcolorbox[auto counter]{finding_rl}[1][]{
  colback=orange!3!white,
  colframe=orange!70!black,
  fonttitle=\tt\small,
  title={\textbf{Summary Of Findings For RL~\thetcbcounter}},
  enhanced,
  boxrule=0.5pt,
  left=1mm,right=1mm,top=1mm,bottom=1mm,
  #1
}

\newtcolorbox[auto counter]{prompt}[2][]{
  colback=gray!10!white,
  colframe=black!75!white,
  fonttitle=\bfseries\small\ttfamily,
  coltitle=white,
  title=#2,
  enhanced,
  boxrule=0.5pt,
  left=1mm,right=1mm,top=1mm,bottom=1mm,
  #1 
}

\title{
\raisebox{-0.25\height}{%
\includegraphics[height=2em]{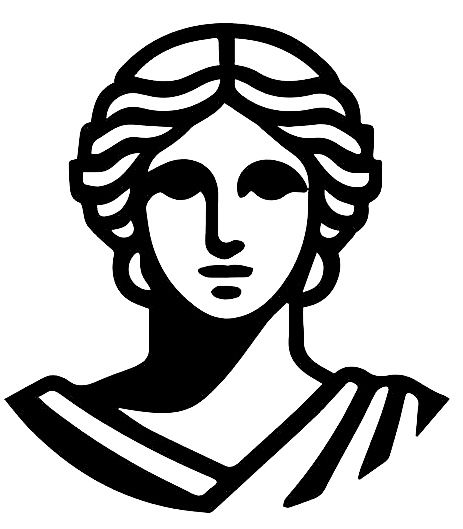}}%
\hspace{0.05em}
\texttt{Aletheia}: What Makes RLVR For Code Verifiers Tick?
}

\author{
Vatsal Venkatkrishna\textsuperscript{\,\insait{}\,\faEnvelope},
Indraneil Paul\textsuperscript{\,\ukp{}}, 
Iryna Gurevych\textsuperscript{\,\insait{}\,\ukp{}}
\newline
\newline
\textsuperscript{\insait{}} INSAIT, Sofia University ``St. Kliment Ohridski'', Bulgaria\\
\textsuperscript{\ukp{}} Ubiquitous Knowledge Processing Lab (UKP Lab), \\
                         Department of Computer Science, Technical University of Darmstadt and \\
                         National Research Center for Applied Cybersecurity ATHENE, Germany \\
\newline
{\raisebox{-0.15\height}{\includegraphics[height=1em]{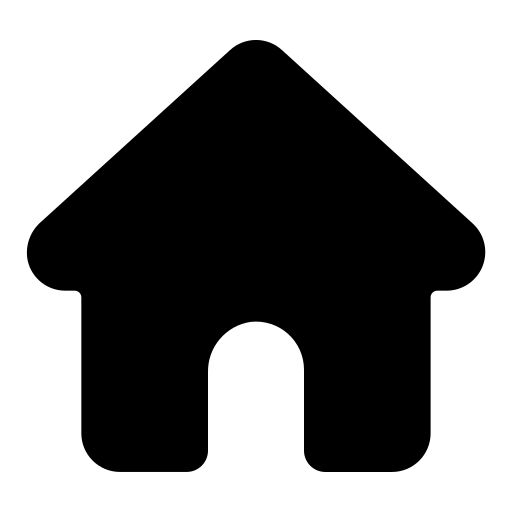}}} \texttt{\href{https://insait-institute.github.io/aletheia/}{insait-institute.github.io/aletheia}} 
}

\def\month{08}  
\def\year{2026} 
\def\openreview{\url{https://openreview.net/forum?id=3rVrBGp0mr}} 

\begin{document}

\maketitle
{
    \def\thefootnote{}\footnotetext{\faEnvelope \hspace{.5em} Corresponding author: \texttt{vatsal.venkatkrishna@insait.ai}}
}

\begin{abstract}
    Multi-domain thinking verifiers trained via Reinforcement Learning with Verifiable Rewards (RLVR) are a cornerstone of modern post-training. However, their adoption in code generation has lagged behind that of execution feedback due to the prohibitive costs of the full RLVR pipeline. In this work, we ablate three primary choices along the performance--cost trade-off in RLVR: intermediate thinking traces, learning from negative samples, and on-policy training. We introduce \textbf{\workname{}}, a controlled, execution-grounded testbed to facilitate a decontaminated analysis of code verifier training recipes across disparate model sizes and covariate shifts across two common verifier application scenarios. Our analysis reveals that the optimal training recipe is scale-dependent: on-policy learning is the primary performance driver for small verifiers, whereas the thinking budget becomes the most vital factor at larger scales. Negative samples play a key role in stabilizing training at large sizes. They have a constant impact on top-1 selection, but are increasingly important for ranking performance as size increases. Our Pareto optimality analysis demonstrates that eliminating on-policy training at larger model scales could yield a verifier that performs comparably to the full RLVR recipe. Furthermore, we find that eschewing thinking traces is a compute-efficient strategy at lower budgets, offering a strong trade-off between training cost and verifier accuracy. We validate our findings across a Best-of-N deployment setting and two external reward model benchmarks, demonstrating that our findings generalize beyond the controlled testbed. Ultimately, our work offers empirical guidance toward training cost-efficient code verifiers and takes a step toward their wider adoption in post-training pipelines for code.
\end{abstract}

\section{Introduction}
\label{sec: intro}
There has been a strong uptick in the adoption of coding assistants that use agentic harnesses such as Claude Code~\citep{claude-code}. Their daily use by software engineers is driven by recent increases in capability via improved post-training~\citep{yang2025qwen3technicalreport,cursor2026composer2technicalreport}. Most post-training recipes require that LLM-generated code be verified against runtime signals~\citep{le2022coderl, shojaee2023executionbasedcodegenerationusing, DBLP:conf/icml/GehringZCMCS25, DBLP:journals/tmlr/LiuZXF00Y23}. However, self-contained executable codes with accompanying test-cases are a scarce resource, even for curated competitive programming datasets, and manual creation does not scale \cite{wang2025codecontests+}. While automatic test-case generation is a common solution~\citep{li2022codecontests, liu2023isyourcode, li2023taco}, it struggles with test coverage and the inherent difficulty of specifying assertions for open-ended tasks. Alternatives like self-contained environment creation~\citep{DBLP:journals/corr/abs-2504-07164, DBLP:journals/corr/abs-2404-00566} and world modelling~\citep{DBLP:journals/corr/abs-2510-02387} can be challenging for compiled languages without mature package managers.

In this work, we revisit surrogate code-execution verifiers~\citep{ni2023lever, li2025codeprm, zeng2025acecoder, DBLP:conf/emnlp/ShiFGZW22, DBLP:conf/icml/ZhangYHLYF023}: models trained to score code snippets based on execution outcomes without actually executing them. In addition to removing the dependence on high-quality test cases, these verifiers obviate code execution and environment-setup overheads in downstream applications like Reinforcement Learning with Verifiable Rewards (RLVR; \citealp{zhu2026codescalerscalingcodellm}) and Best-of-N (BoN) inference~\citep{zeng2025acecoder}. Such verifiers can additionally leverage the code understanding and generalization capabilities of LLMs to provide granular feedback signals for long-horizon tasks where sparse episodic rewards may fail to adequately guide convergence~\citep{DBLP:journals/corr/abs-2502-01456}.

\begin{figure*}[!t]
    \centering
    \includegraphics[width=1\linewidth]{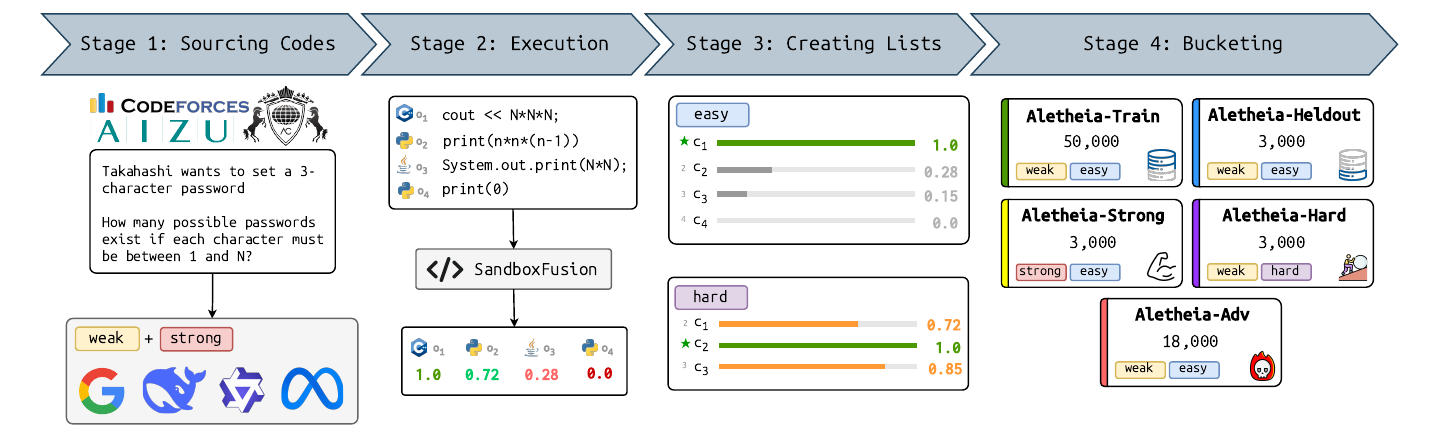}
    \caption{\textbf{\workname{} testbed curation pipeline.} We follow a four-stage procedure: (1) generating solutions for competition-level programming questions from \texttt{CodeContests$^{\text{+}}$}~\citep{wang2025codecontests+} from a pool of \weak{} and \sstrong{} open-source LLMs; (2) obtaining ground-truth pass rates (\texttt{PRs}) for the obtained codes through execution using SandboxFusion; (3) constructing lists of \texttt{2-5} candidates where exactly one is fully correct. These lists are either \easy{} (second best \texttt{PR$\,<\,$0.5}) or \hard{} (second best \texttt{PR$\,\in\,$[0.7, 0.9]}); and (4) partitioning the resulting data into completely disjoint training and evaluation sets across three covariate shifts: strong generators, hard comparisons, and adversarial prompts. Refer to \cref{sec: data-creation} for a detailed description.}
    \label{fig: dset-curation}
\end{figure*}

\noindent\textbf{The lack of RLVR adoption for code verifiers.} Recently, generative verifiers have proliferated in post-training pipelines for reasoning-heavy domains such as math and science~\citep{liu-etal-2025-compassverifier, DBLP:journals/corr/abs-2505-14652, DBLP:journals/corr/abs-2510-06499}. Often, these verifiers are trained to reason before answering using RLVR~\citep{chenRMR1RewardModeling2025, DBLP:journals/corr/abs-2505-14268}, which boosts interpretability~\citep{DBLP:journals/corr/abs-2507-17746} and mitigates reward hacking~\citep{DBLP:journals/corr/abs-2504-13914}. However, these advances have made fewer inroads into code verifiers, which are predominantly encoder-only regression models~\citep{zhu2026codescalerscalingcodellm, DBLP:conf/icml/ZhangYHLYF023, DBLP:conf/emnlp/ShiFGZW22}. Although such models can approximate code execution outcomes~\citep{DBLP:journals/corr/abs-2509-26476}, they remain brittle against adversarial and semantic-preserving transformations even at scale~\citep{DBLP:journals/corr/abs-2504-04372, DBLP:journals/corr/abs-2502-11167}, pointing to the need for richer, generative approaches. Yet the lack of adoption is unsurprising: the full RLVR recipe is expensive and demands intricate orchestration of rollout, behavior, and reference policies. Thus, we study verifier training along three axes where RLVR differs from cheaper post-training recipes, namely: generating long intermediate reasoning traces (\textbf{Thinking}), learning from both positive and negative samples (\textbf{Negatives}), and learning from data generated by an updated policy (\textbf{Online}).


Although these RLVR components have been studied independently for generator training~\citep{tajwar2024preference, lanchantinBridgingOfflineOnline2025, zhu2025surprising}, no equivalent analysis exists for verifiers. This gap is significant due to the well-documented inconsistencies between the generative and verification abilities of LLMs~\citep{DBLP:journals/corr/abs-2310-01846,rodriguez2025rankalign,song2025mind}. Unlike generators, evaluating verifiers requires going beyond downstream task accuracy. \citet{razin2026what} show that even a perfectly accurate verifier can induce a flat loss landscape and stall learning. Moreover, LLM verifiers are known to break down under distribution shifts~\citep{eisenstein2024helping} and adversarial prompting~\citep{moonDontJudgeCode2025,lamCodeCrashStressTesting2025}, which can corrupt the generator being supervised. The optimal code verifier training recipe also likely varies with scale~\citep{DBLP:journals/corr/abs-2001-08361,Hoffmann2022TrainingComputeOptimal} and depends on the application, such as BoN selection or use as an RL reward model~\citep{wenRethinkingRewardModel2024,kimRethinkingRewardModel2025}.

\noindent\textbf{The need for an evaluation testbed for verifiers.} Although integrating verifiers into their downstream use cases seems like a natural way to evaluate verifiers, it is noisy and the resulting policy improvement is a poor proxy for reward quality. This is especially true for verifiers used as RL reward models. Such methods are known to succeed only when the base model learns certain task-specific primitives during earlier training phases~\citep{setlur2026e,DBLP:journals/corr/abs-2512-07783,DBLP:journals/corr/abs-2506-20512}, making downstream performance an amalgam of prior capability, reward signal, and the optimization process~\citep{wenRethinkingRewardModel2024,kimRethinkingRewardModel2025}. These confounds are so stark that certain models improve even under random rewards~\citep{shao2025spuriousrewardsrethinkingtraining}.

Additionally, the pipeline is expensive and slow to iterate on, which significantly hurts reproducibility and slows progress~\citep{DBLP:conf/iclr/FrickLCCAJZGS25}. Evidence suggests that the early steps in RLVR merely improve sampling efficiency~\citep{yueDoesReinforcementLearning2025,wu2025invisible} and true expansion of reasoning capabilities requires prolonged training~\citep{liuProRLProlongedReinforcement2025,DBLP:journals/corr/abs-2510-04028}, which puts a principled evaluation out of reach for most reasonable budgets. Even recent RL scaling studies require discarding the first ${\sim}$\texttt{1.5k} GPU-hours of every run~\citep{devvrit2026the}, much like pre-training scaling studies~\citep{li2025misfitting,porian2024resolving}.
To sidestep these limitations, we create \textbf{\workname{}}, an execution-grounded testbed for evaluating verifier training recipes. We mirror downstream evaluation scenarios through a decontaminated testbed that enforces a strict training\,--\,evaluation partition, isolating algorithmic robustness to out-of-distribution (OOD) scenarios from simple data exposure. We ablate the three RLVR components mentioned earlier across three covariate shifts commonly encountered in downstream evaluations: stronger generators than the ones seen during training (\rqone{}), incorrect solutions that are semantically very close to the correct one (\rqtwo{}), and adversarially modified code snippets (\rqfour{}). We validate our findings across three model sizes (\texttt{1.5/7/14B}), uncovering scale-dependent training dynamics. We construct our testbed in accordance with prior work on reward model evaluations~\citep{fengAreWeRight2025,wenRethinkingRewardModel2024,kimRethinkingRewardModel2025} and evaluate each verifier's ability to select the best candidate (top-1 selection) and reproduce the full ranked order (reranking) as a proxy for BoN and RL performance respectively. These metrics are evaluated over lists of \texttt{2\,--\,5} codes. Since these metrics are imperfect proxies, we additionally validate our findings across a downstream BoN selection task and two external benchmarks.


Our analysis reveals that while RLVR is the best-performing method to train verifiers in most evaluation settings, the contribution of each ablated component varies with scale. Across both downstream application scenarios, we find that on-policy learning is critical for small verifiers, but its contribution diminishes as model size increases. Conversely, thinking traces offer limited benefits at smaller scales but become essential for \texttt{14B} models. Meanwhile, negative samples provide a near-consistent boost to all sizes for top-1 selection but a monotonic boost for reranking, and play a critical role in stabilizing training at larger scales. Utilizing additional compute for verifiers at inference time using self-consistency yields modest gains and cannot compensate for any of the core components we analyze.

Additionally, we study the Pareto optimality of each ablated recipe with respect to cost and performance. \texttt{DPO-Think-14B} is competitive with the full \texttt{GRPO-Think} recipe at \texttt{5.2$\times$} lower cost, and is the only 14B method on the Pareto frontier across all evaluations simultaneously. The full GRPO recipe is warranted only when peak performance across generator shifts and adversarial perturbations is required. For \rqtwo{}, practitioners should favor \texttt{DPO-Think-14B} for top-1 selection or \texttt{RAFT-14B} for reranking, as on-policy training provides no cost-adjusted benefit. \texttt{GRPO-Instruct-7B} is a good low-budget option on the Pareto frontier for all evaluations, but has a weaker absolute performance than \texttt{DPO-Think-14B}.
We summarize our contributions as follows.

\begin{itemize}[leftmargin=*,noitemsep]
    \vspace{-.5em}
    \item We introduce \textbf{\workname}, an execution-grounded testbed that enables a decontaminated evaluation across three covariate shifts: stronger generators, harder comparisons, and adversarial responses (\cref{sec: setup}).
    \item We ablate the contributions of three core RLVR components to code verifier performance across different model sizes (\texttt{1.5/7/14B}): thinking traces, on-policy learning, and negative samples (\cref{sec: results}).
    \item We conduct a cost-benefit analysis of the RLVR training recipe for robust code verifiers (\cref{sec: analysis}).
    \item We validate our findings across a downstream Best-of-N selection and two external benchmarks (\cref{sec: downstream}).
\end{itemize}

\section{Experimental Setup}
\label{sec: setup}
\subsection{Testbed Creation}
\label{sec: data-creation}

A code verifier's value is ultimately determined by its downstream use, either in BoN inference or as a reward model inside an RLVR pipeline. Evaluating verifiers against either of these uses, however, is far from straightforward. The de-facto offline metric, paired accuracy on RewardBench-style benchmarks~\citep{lambert-etal-2025-rewardbench,liuRMBenchBenchmarkingReward2024,tanJudgeBenchBenchmarkEvaluating2025}, is cheap but only weakly predictive of downstream performance: it is insensitive to score separation across the full candidate list~\citep{razin2026what}, to discrimination under near-correct distractors~\citep{kimRethinkingRewardModel2025,fengAreWeRight2025}, and to the distributional coverage required as the policy drifts from the verifier's training distribution~\citep{eisenstein2024helping,wenRethinkingRewardModel2024}.

To reliably evaluate verifier training recipes, we construct \textbf{\workname{}}: a controlled,  decontaminated, execution-grounded testbed. Each prompt is a listwise selection problem, which disincentivizes models from gaming rewards via random guessing.
Following prior work on reliable verifier evaluation~\citep{kimRethinkingRewardModel2025}, we source codes from a wide range of model families and sizes (\cref{tab: generator-models}), but maintain a single generator within a single prompt to mimic downstream evaluation~\citep{fengAreWeRight2025}. Concretely, given a coding problem $\rvx$ and $N$ distinct candidate solutions $\rmC = \{\rvc_n\}_{n=1}^N$ with execution-derived pass rates $\rmP = \{\rp_n\}_{n=1}^N$, the verifier generates $\rvy = (\rvz, \rvo)$: a reasoning trace $\rvz$ followed by a predicted index $\rvo \in \{1, \ldots, N\}$ identifying the best candidate. By construction, exactly one candidate is fully correct ($\max_n \rp_n = 1$), so the ground truth is unambiguous and the reward $\1[\rvo = \argmax_n \rp_n]$ is trivially computable.

We source competitive programming problems from \texttt{CodeContests$^+$}~\citep{wang2025codecontests+}, each with \texttt{$\approx$\,25} synthetic test cases. We only include high-quality problems with true positive rates and true negative rates \texttt{>\,\,0.9} against a pre-evaluated user-submitted solution pool, and discard samples with \texttt{<\,\,5} test cases or a time limit \texttt{>\,\,3} seconds. This gives us \texttt{4903} programming questions. We generate \texttt{50} completions for each problem in \texttt{Python}, \texttt{C++}, and \texttt{Java} at a sampling temperature of \texttt{1.0} using a mix of \weak{} and \sstrong{} LLMs (\cref{tab: generator-models}).
The resulting code snippets are executed using \texttt{SandboxFusion}\footnote{\href{https://github.com/bytedance/SandboxFusion}{\scalerel*{\includegraphics{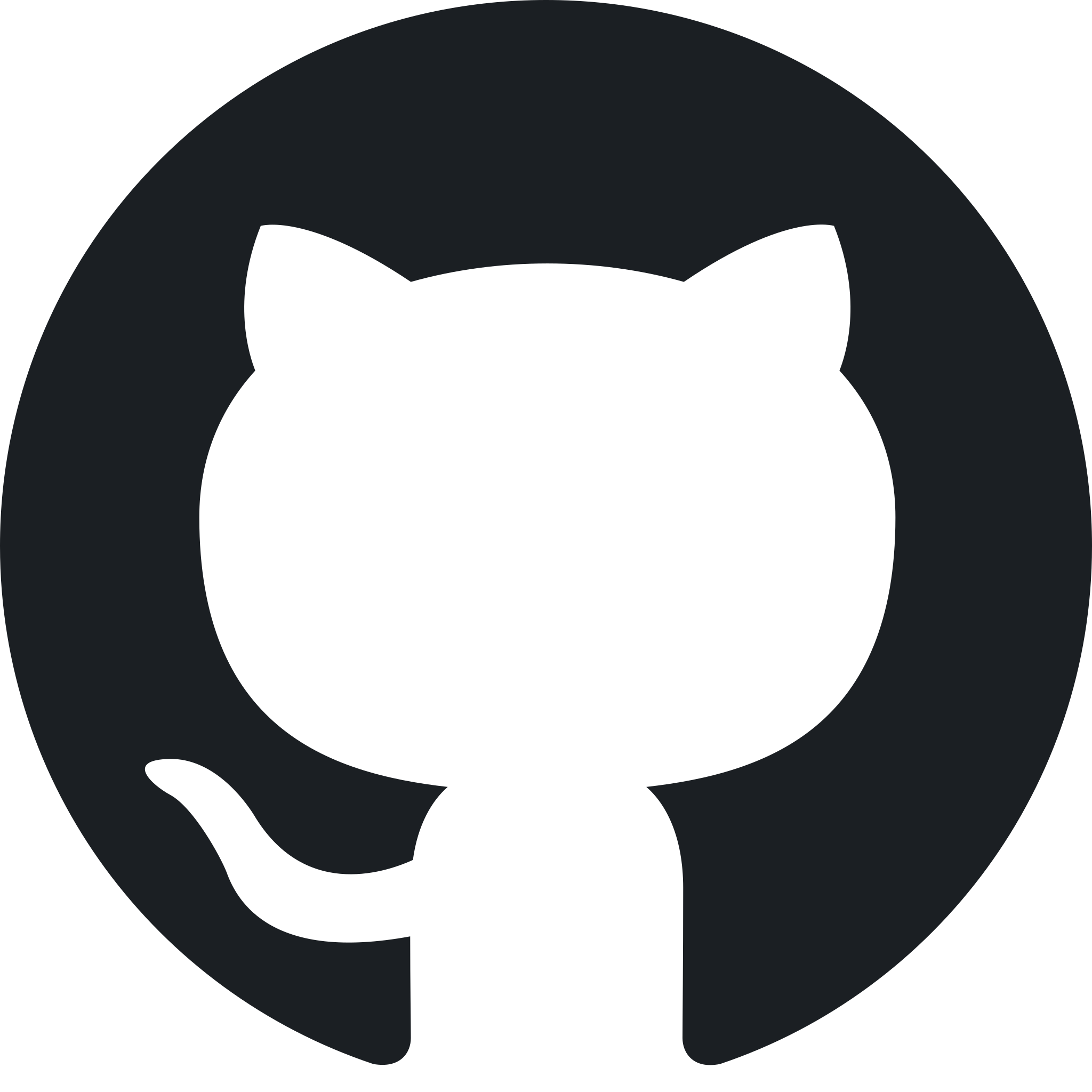}}{\textbf{C}} \texttt{bytedance/SandboxFusion}}} to calculate their pass rates (PRs); the percentage of test cases passed for each code. We construct lists of \texttt{2\,--\,5} solution codes with distinct PRs. We divide these lists into \easy{} and \hard{} buckets based on their PRs; \easy{} contains lists where \texttt{PR$_{\text{incorrect}}\!\,\in$\,[0, 0.5]} and \hard{} has \texttt{PR$_{\text{incorrect}}\!\,\in$\,[0.7, 0.9]}, making them harder to distinguish from the correct code.

We subsample the \weak{}-\easy{} bucket to \texttt{50,000} instances, dubbed \textbf{\train{}}. Additionally, we create four evaluation datasets, evenly distributed by list length and programming language:
\begin{itemize}[leftmargin=*,noitemsep]
    \item \textbf{\heldout{}.} An in-distribution evaluation set containing \easy{} comparisons by \weak{} models.
    \item \textbf{\rqone{}.} \easy{} comparisons by \sstrong{} models, which tests the verifier's robustness to a shift in the generator's capability, without altering the quality of the codes being compared~\citep{DBLP:journals/corr/abs-2509-17995}.
    \item \textbf{\rqtwo{}.} \hard{} comparisons generated by \weak{} models. Verifier performance on this dataset indicates their ability to generalize from easy to hard~\citep{haseUnreasonableEffectivenessEasy2024, sunEasytoHardGeneralizationScalable2024}.
    \item \textbf{\rqfour{}.} It evaluates the adversarial robustness of our verifiers. We apply three positive and negative modifications to the incorrect and correct codes in \heldout{} respectively (see \cref{sec: adversarial-mods}), based on prior work on biases in LLM judges~\citep{lamCodeCrashStressTesting2025, hwangCanYouTrick2025,moonDontJudgeCode2025}.
\end{itemize}
All OOD evaluations introduce a unidirectional shift from the training data. We further reduce contamination by ensuring no overlap between training and evaluation coding problems.
The prompts used in this work are listed in \cref{sec: prompt_templates}. We summarize dataset statistics in \cref{tab: dset-stats}. We calculate the average code similarities in our datasets by encoding them using \texttt{Qwen3-Embedding-8B}~\citep{qwen3embedding}, which achieves stellar performance on text embedding benchmarks like MTEB~\citep{muennighoff2022mteb, enevoldsen2025mmtebmassivemultilingualtext}. The codes within a single prompt are quite similar by virtue of being generated from the same model, with all similarity scores being \texttt{>\,\,0.88}. The codes in \rqtwo{} particularly stand out, with an average similarity of \texttt{0.93} and a smaller average length of \texttt{186} tokens. Since all codes in this dataset pass at least \texttt{70\%} of the test cases, they are semantically very close to the correct solution and thus harder to distinguish.

\begin{table}[!t]
    \begin{minipage}[c]{0.48\linewidth}
        \centering
        \caption{\textbf{Descriptive statistics for our datasets.} Code lengths are measured in tokens and averaged over the lists within a single instance. Average similarities are computed as the cosine similarities between the embeddings of the codes in each list.}
        \label{tab: dset-stats}
        \tt
        \small
        \scalebox{0.8}{
    \begin{tabular}{l|cccc}
        \toprule
        \textbf{Name}       & \textbf{\#Rows} & \textbf{\#Qtns} & \textbf{Code Len.} & \textbf{Code Sim.} \\\midrule
        \textbf{\train{}}   & 50,000          & 1247            & 208                & 0.896              \\
        \textbf{\heldout{}} & 3,000           & 456             & 236                & 0.893              \\
        \textbf{\rqone{}}   & 3,000           & 151& 287                & 0.898              \\
        \textbf{\rqtwo{}}   & 3,000           & 137             & 186                & 0.931              \\
        \textbf{\rqfour{}}  & 18,000          & 456             & 246                & 0.882              \\\bottomrule
    \end{tabular}
}

    \end{minipage}
    \hfill
    \begin{minipage}[c]{0.48\linewidth}
        \centering
        \caption{\textbf{Generators used for our datasets.} \texttt{$\Delta$Score} is the difference in \texttt{BigCodeBench-Instruct} scores~\citep{zhuo2024bigcodebench} between the \sstrong{} and \weak{} models, representing the gap in their abilities. We use the \texttt{-Instruct} variants for all the generators.}
        \label{tab: generator-models}
        \tt
        \small
        \scalebox{0.8}{
    \renewcommand{\arraystretch}{1.16}
    \begin{tabular}{lccc}
        \toprule
        \textbf{Model Family}                                                                                                            & \textbf{Weak} & \textbf{Strong} & \textbf{$\Delta$Score} \\
        \midrule
        \raisebox{-0.1 \height}{\includegraphics[height=0.8em]{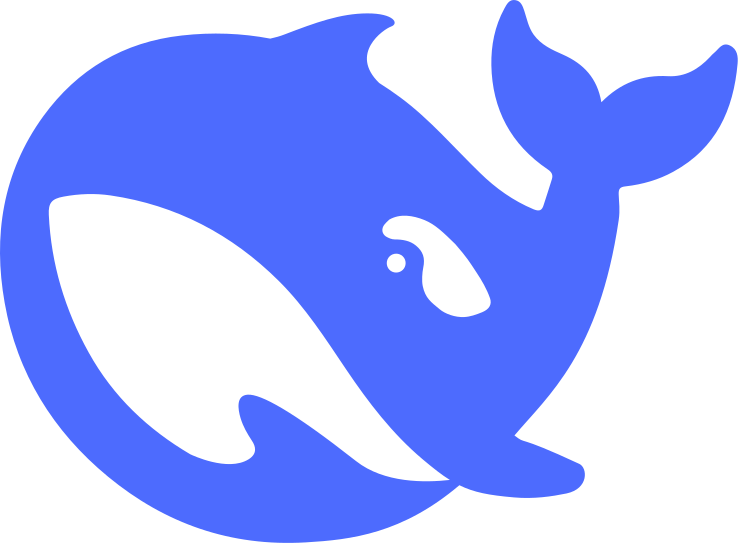}}\hspace{-0.1em} \texttt{deepseek-ai/deepseek-coder} & 6.7B          & 33B             & 6.5                    \\
        \raisebox{-0.1 \height}{\includegraphics[height=0.8em]{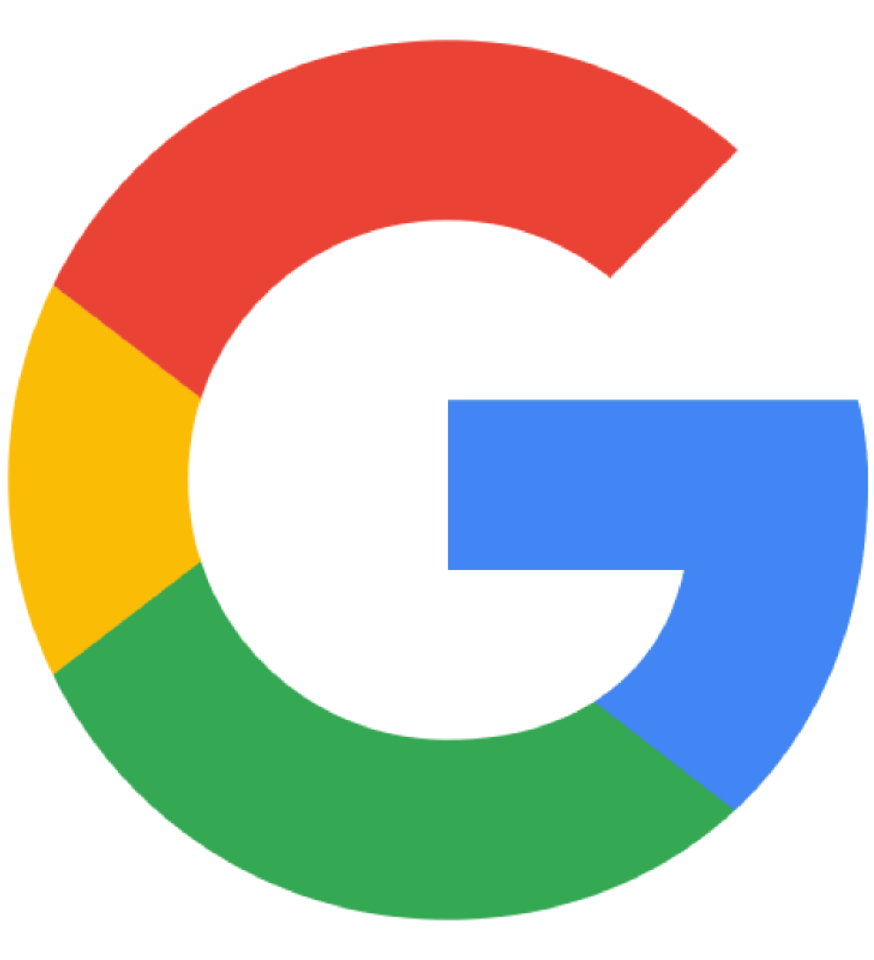}}\hspace{0.2em} \texttt{google/gemma2}                 & 9B            & 27B             & 5.1                    \\
        \raisebox{-0.1 \height}{\includegraphics[height=0.8em]{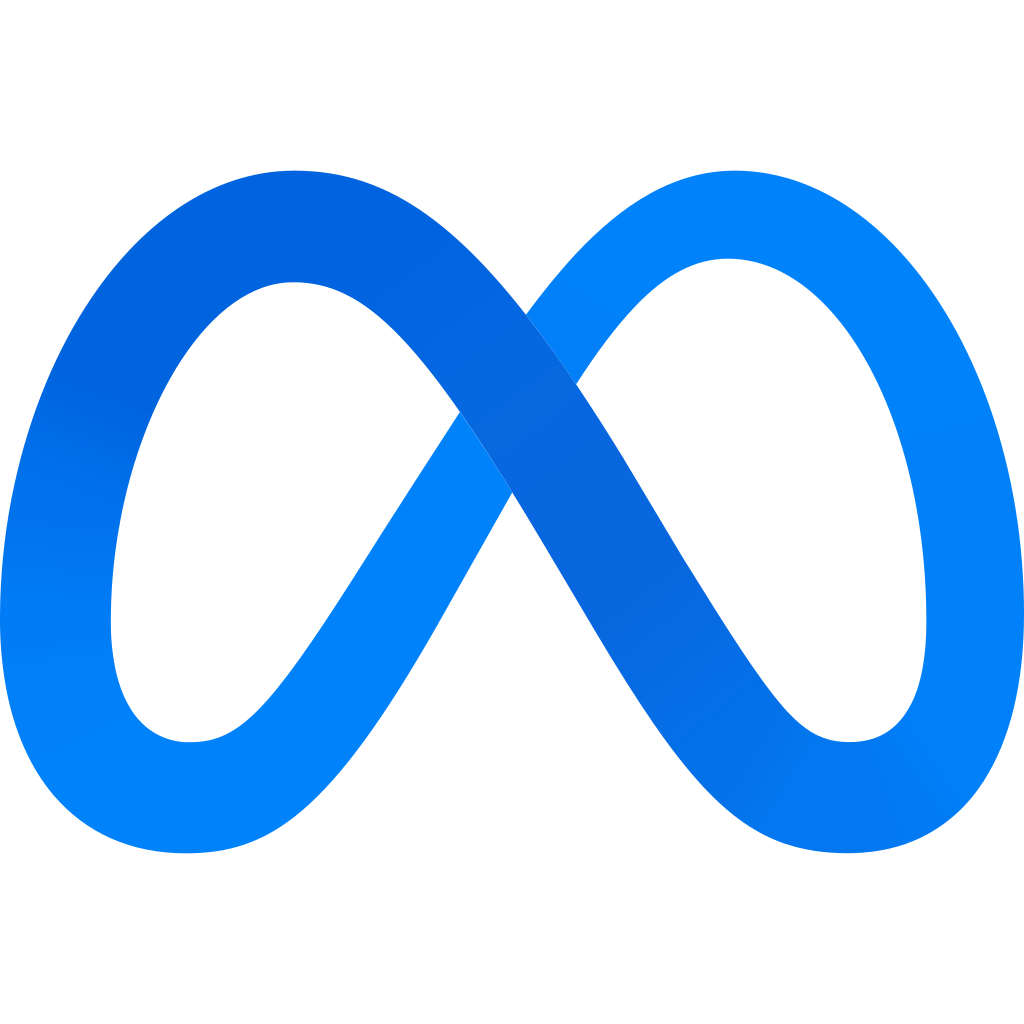}}\hspace{0.2em} \texttt{meta/llama-3.1}                  & 8B            & 70B             & 13.3                   \\
        \raisebox{-0.1 \height}{\includegraphics[height=0.8em]{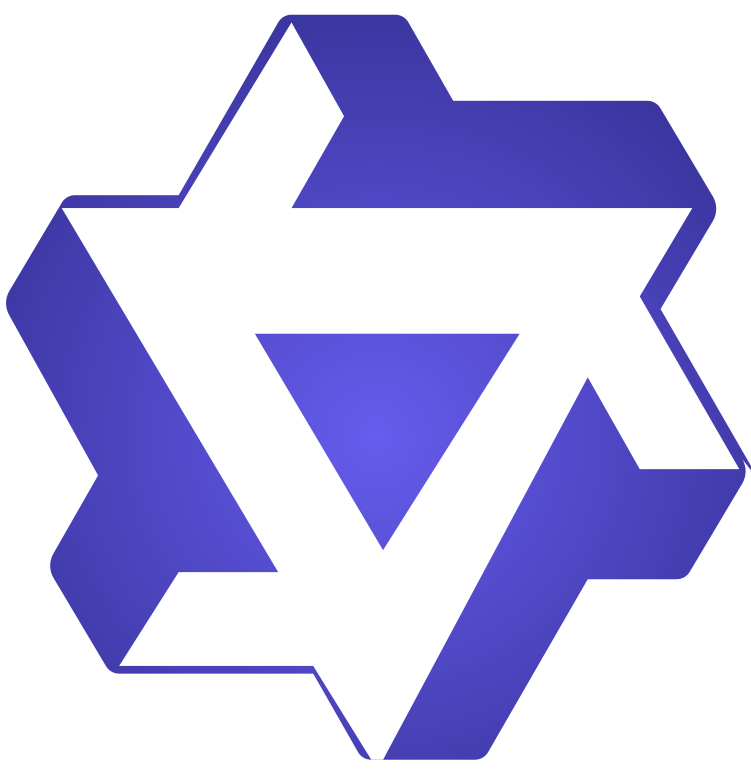}}\hspace{0.2em} \texttt{qwen/qwen2.5-coder}              & 7B            & 32B             & 8.6                    \\
        \bottomrule
    \end{tabular}
}
    \end{minipage}
\end{table}

\subsection{Evaluation Metrics}
\label{sec: metrics}
The \workname{} testbed is carefully designed to evaluate reward models in their two most common downstream usecases via complementary metrics. Crucially, our task formulation matches BoN inference almost exactly: selecting the $\text{PR}\!=\!1.0$ winner from a generator-matched pool across a variable number of candidates. Moreover, recent work suggests that accuracy evaluated over multiple comparisons and across a range of response quality predicts BoN performance better than paired accuracy~\citep{wenRethinkingRewardModel2024}, although this is still a weak positive correlation. Thus, we report the average top-1 selection accuracy (\texttt{ListAcc}) across all evaluation datasets as a predictor of downstream BoN performance. During evaluation, we generate responses using a temperature of \texttt{0.6} and \texttt{top-p\,=\,0.95} nucleus sampling. For each prompt we draw \texttt{16} generations to obtain a low-variance estimate of the per-prompt score. Each reported  metric is  \texttt{$\bar{\text{x}}_{\text{1.96}\,\sigma/\sqrt{\text{N}}}$}, where \texttt{$\bar{\texttt{x}}$} is the mean of these scores over the \texttt{N} prompts in the evaluation dataset, and the subscript is the \texttt{95\%} confidence interval on that mean.

However, accuracy alone is often insufficient to predict the utility of a verifier as an RL reward-model~\citep{wenRethinkingRewardModel2024,fengAreWeRight2025}. As mentioned in \cref{sec: intro}, direct RL integration may also not be indicative of reward quality and is prohibitively expensive. Thus, we evaluate our verifiers' ability to reconstruct the full ranked order of $N$ candidates rather than selecting the best candidate, which may better predict downstream RL performance~\citep{wenRethinkingRewardModel2024}.

We adopt Kendall's $\tau$-b (\ktau{}) as our reranking metric. For each list of $N$ candidates, we issue all $\binom{N}{2}$ pairwise comparisons as independent verifier generations and parse each verdict to identify the winning candidate. We score every candidate $\rvc_n$ by its number of pairwise wins and compute
\begin{equation*}
    \ktau{} = \frac{n_C - n_D}{\sqrt{(n_C + n_D + T_w)(n_C + n_D + T_p)}},
\end{equation*}
where $n_C$ and $n_D$ are the numbers of candidate pairs ranked concordantly and discordantly by the predicted win counts against the execution-derived pass rates $\rmP$, and $T_w$, $T_p$ count pairs tied in the predicted win counts and in $\rmP$ respectively ($T_p=0$ in our case). We use the tie-corrected $\tau$-b rather than $\tau$-a because candidates routinely tie in predicted win counts. We report \ktau{} averaged over lists in each evaluation set. A \texttt{\ktau{}\!\,\,=\,\,\!+1} implies a perfectly ordered list, \texttt{0} an uninformative ordering, and \texttt{-1} indicates a fully reversed one. Unlike top-1 selection accuracy, \ktau{} uses every pair in the candidate list and remains discriminative even when correct and incorrect codes have near-identical pass rates, which captures the score-separation property required by downstream policy-gradient training~\citep{razin2026what}.

\subsection{Training Details}
We validate our findings across a wide range of model sizes and training parameter counts, including \texttt{1.5B}, \texttt{7B}, and \texttt{14B} for each method. Unless explicitly mentioned, we initialize each method from the \texttt{DeepSeek-R1-Distill-Qwen2.5} models~\citep{deepseekai2025deepseekr1incentivizingreasoningcapability} because they have been warm-started to generate reasoning traces before answering. To ensure a fair comparison, all methods are trained for an identical number of gradient updates. For on-policy methods, we generate \texttt{16} responses at a high sampling temperature of \texttt{1.0} and award a \texttt{+1} to generations that identify the correct candidate, and \texttt{0} otherwise. We also apply a \texttt{-1} penalty to generations that violate the format. We provide a detailed description of our training setup in \cref{sec: detailed-hparams} and experiment with alternate reward formulations in \cref{sec: alt-rewards}.

\section{Research Questions and Results}
\label{sec: results}
Our analysis employs a series of controlled experiments to analyze the contributions of Thinking, Negatives, and Online components. We use \texttt{GRPO}~\citep{shaoDeepSeekMathPushingLimits2024} as a baseline that includes all three components and compare it with algorithms that lack the single component under study. This design choice minimizes confounding factors to isolate each component's individual contribution as much as possible. We study the effect of removing two components in \cref{sec: sft-results}.

\subsection{\texttt{RQ1}: Do Code Verifiers Need to Generate Long Reasoning Traces?}
\label{sec: think-toggle}
\begin{table*}[!t]
    \centering
    \caption{\textbf{Results for ablating thinking trace generation (\raisebox{-0.2em}{\includegraphics[height=1em]{Figures/thinking.png}}).} We report list accuracy and Kendall $\tau$ scores for BoN and RL, respectively. For both metrics, thinking-style traces offer little benefit to small models but are essential for larger models. Increasing the reasoning budget to 16k is almost always useful, but the style of traces alone is most impactful for larger models. Thinking is especially critical for \rqtwo{}. Subscripts are 95\% confidence intervals.}
    \small
    \tt
    \setlength{\tabcolsep}{2.8pt}
    \scalebox{0.8}{

        \begin{tabular}{clcc|w{c}{1.3cm}w{c}{1.3cm}|w{c}{1.3cm}w{c}{1.3cm}|w{c}{1.3cm}w{c}{1.3cm}|w{c}{1.3cm}w{c}{1.3cm}|w{c}{1.1cm}w{c}{1.1cm}}
            \multicolumn{14}{c}{\thinking{}~\texttt{\textbf{Thinking}} \quad \negatives{}~\texttt{\textbf{Negatives}} \quad \online{}~\texttt{\textbf{Online}}}                                                                                                                                                                                                                                                                                                                                                 \\
            \toprule
                                                   & \multirow{2}{*}{\textbf{Method}} & \multirow{2}{*}{\textbf{Size}} & \multirow{2}{*}{\textbf{Cost (\$)}} & \multicolumn{2}{c|}{\textbf{\heldout{}}} & \multicolumn{2}{c|}{\textbf{\rqone{}}} & \multicolumn{2}{c|}{\textbf{\rqtwo{}}} & \multicolumn{2}{c|}{\textbf{\rqfour{}}} & \multicolumn{2}{c}{\makebox[1em][c]{\textbf{Average}}}                                                                                                                    \\
            \cmidrule(lr){5-6} \cmidrule(lr){7-8} \cmidrule(lr){9-10} \cmidrule(lr){11-12} \cmidrule(lr){13-14}
                                                   &                                  &                                &                                     & \texttt{BoN}                             & \texttt{RL}                            & \texttt{BoN}                           & \texttt{RL}                             & \texttt{BoN}                                           & \texttt{RL}          & \texttt{BoN}         & \texttt{RL}          & \texttt{BoN}         & \texttt{RL}          \\ \midrule
            -                                      & Random                           &                                & -                                   & \scval{32.08}                            & \rcval{0.00}                           & \scval{32.08}                          & \rcval{0.00}                            & \scval{32.08}                                          & \rcval{0.00}         & \scval{32.08}        & \rcval{0.00}         & \scval{32.08}        & \rcval{0.00}         \\ \midrule
            \nothinking{}\,\negatives{}\,\online{} & GRPO-Instruct                    & \multirow{4}{*}{1.5B}          & 0.587                               & \scvalc{38.78}{0.85}                     & \rcvalc{11.90}{2.18}                   & \scvalc{40.51}{0.85}                   & \rcvalc{13.09}{2.29}                    & \scvalc{31.22}{0.81}                                   & \rcvalc{4.11}{2.37}  & \scvalc{30.99}{0.33} & \rcvalc{2.16}{2.24}  & \scvalc{35.41}{0.71} & \rcvalc{7.82}{2.27}  \\
            \thinking{}\,\negatives{}\,\online{}   & GRPO-Think-4k                    &                                & 1.208                               & \scvalc{42.73}{0.64}                     & \rcvalc{18.48}{2.38}                   & \scvalc{40.70}{0.61}                   & \rcvalc{13.08}{2.38}                    & \scvalc{36.32}{0.57}                                   & \rcvalc{5.36}{2.40}  & \scvalc{31.15}{0.25} & \rcvalc{2.32}{2.44}  & \scvalc{37.76}{0.52} & \rcvalc{9.81}{2.40}  \\
            \thinking{}\,\negatives{}\,\online{}   & GRPO-Think-8k                    &                                & 2.491                               & \scvalc{46.82}{0.62}                     & \rcvalc{22.90}{2.39}                   & \scvalc{43.65}{0.59}                   & \rcvalc{17.56}{2.40}                    & \scvalc{41.61}{0.61}                                   & \rcvalc{10.41}{2.41} & \scvalc{38.09}{0.25} & \rcvalc{9.50}{2.54}  & \scvalc{42.55}{0.52} & \rcvalc{15.09}{2.43} \\
            \thinking{}\,\negatives{}\,\online{}   & GRPO-Think-16k                   &                                & 7.806                               & \scvalc{49.58}{0.65}                     & \rcvalc{27.49}{2.32}                   & \scvalc{46.09}{0.63}                   & \rcvalc{23.57}{2.33}                    & \scvalc{40.74}{0.64}                                   & \rcvalc{8.92}{2.48}  & \scvalc{40.97}{0.26} & \rcvalc{16.45}{2.53} & \scvalc{44.38}{0.55} & \rcvalc{19.11}{2.42} \\ \midrule
            \nothinking{}\,\negatives{}\,\online{} & GRPO-Instruct                    & \multirow{4}{*}{7B}            & 2.069                               & \scvalc{57.74}{0.88}                     & \rcvalc{35.42}{2.00}                   & \scvalc{51.80}{0.89}                   & \rcvalc{29.08}{2.35}                    & \scvalc{38.59}{0.87}                                   & \rcvalc{9.88}{2.40}  & \scvalc{52.20}{0.37} & \rcvalc{29.20}{2.18} & \scvalc{50.07}{0.75} & \rcvalc{25.90}{2.23} \\
            \thinking{}\,\negatives{}\,\online{}   & GRPO-Think-4k                    &                                & 3.561                               & \scvalc{59.54}{0.71}                     & \rcvalc{37.40}{2.25}                   & \scvalc{55.00}{0.70}                   & \rcvalc{32.03}{2.34}                    & \scvalc{46.73}{0.74}                                   & \rcvalc{14.92}{2.40} & \scvalc{44.04}{0.30} & \rcvalc{21.84}{2.50} & \scvalc{51.32}{0.61} & \rcvalc{26.55}{2.37} \\
            \thinking{}\,\negatives{}\,\online{}   & GRPO-Think-8k                    &                                & 7.179                               & \scvalc{65.03}{0.57}                     & \rcvalc{35.22}{2.09}                   & \scvalc{56.96}{0.57}                   & \rcvalc{32.40}{2.41}                    & \scvalc{53.16}{0.65}                                   & \rcvalc{17.81}{2.40} & \scvalc{52.03}{0.26} & \rcvalc{28.15}{2.35} & \scvalc{56.76}{0.51} & \rcvalc{28.40}{2.31} \\
            \thinking{}\,\negatives{}\,\online{}   & GRPO-Think-16k                   &                                & 15.101                              & \scvalc{74.81}{0.57}                     & \rcvalc{51.52}{1.73}                   & \scvalc{67.28}{0.60}                   & \rcvalc{46.30}{2.03}                    & \scvalc{53.11}{0.69}                                   & \rcvalc{20.31}{2.44} & \scvalc{65.04}{0.26} & \rcvalc{46.16}{2.14} & \scvalc{65.05}{0.53} & \rcvalc{41.07}{2.08} \\ \midrule
            \nothinking{}\,\negatives{}\,\online{} & GRPO-Instruct                    & \multirow{4}{*}{14B}           & 9.463                               & \scvalc{63.45}{0.82}                     & \rcvalc{41.70}{1.98}                   & \scvalc{55.11}{0.84}                   & \rcvalc{31.01}{2.17}                    & \scvalc{44.15}{0.84}                                   & \rcvalc{16.97}{2.32} & \scvalc{54.24}{0.35} & \rcvalc{27.51}{2.19} & \scvalc{54.26}{0.71} & \rcvalc{29.30}{2.17} \\
            \thinking{}\,\negatives{}\,\online{}   & GRPO-Think-4k                    &                                & 8.558                               & \scvalc{73.23}{0.67}                     & \rcvalc{46.14}{1.92}                   & \scvalc{64.95}{0.70}                   & \rcvalc{39.74}{2.01}                    & \scvalc{54.56}{0.77}                                   & \rcvalc{21.34}{2.31} & \scvalc{58.09}{0.31} & \rcvalc{31.46}{2.27} & \scvalc{62.69}{0.61} & \rcvalc{34.67}{2.13} \\
            \thinking{}\,\negatives{}\,\online{}   & GRPO-Think-8k                    &                                & 14.900                              & \scvalc{78.37}{0.60}                     & \rcvalc{50.23}{1.86}                   & \scvalc{69.87}{0.65}                   & \rcvalc{41.73}{1.98}                    & \scvalc{61.74}{0.73}                                   & \rcvalc{27.45}{2.30} & \scvalc{65.71}{0.29} & \rcvalc{39.39}{2.12} & \scvalc{68.91}{0.57} & \rcvalc{39.70}{2.06} \\
            \thinking{}\,\negatives{}\,\online{}   & GRPO-Think-16k                   &                                & 36.992                              & \scvalc{88.02}{0.45}                     & \rcvalc{60.10}{1.64}                   & \scvalc{83.65}{0.49}                   & \rcvalc{60.21}{1.74}                    & \scvalc{66.84}{0.70}                                   & \rcvalc{30.95}{2.26} & \scvalc{83.67}{0.21} & \rcvalc{56.84}{1.63} & \scvalc{80.54}{0.46} & \rcvalc{52.03}{1.82} \\ \bottomrule
        \end{tabular}
    }
    \label{tab: think-toggle-merged}
\end{table*}

\begin{finding_bon}
    \scriptsize
    \begin{itemize}[leftmargin=*, noitemsep]
        \item \texttt{The contribution of thinking traces to verifier performance increases monotonically with model scale.}
        \item \texttt{Training reasoning budget follows a similar pattern: 1.5B saturates beyond 8k, but 7-14B continue improving up to 16k.}
        \item \texttt{Thinking traces are crucial for \easy{}-to-\hard{} generalization.}
        \item \texttt{Reasoning-style traces enable models to utilize increased inference compute.}
    \end{itemize}
\end{finding_bon}
\begin{finding_rl}
    \scriptsize
    \begin{itemize}[leftmargin=*, noitemsep]
        \item \texttt{The contribution of thinking traces to RL reward model performance similarly increases with model size.}
        \item \texttt{Training reasoning budget yields continued gains in \ktau{} up to 16k for all verifier sizes.}
        \item \texttt{\ktau{} and ListAcc agree on general trends, but the magnitude of each effect varies.}
    \end{itemize}
\end{finding_rl}

\paragraph{Background.} Thinking traces significantly boost LLM performance~\citep{wei2022chain, kojima2022large}, but the source of these gains is ambiguous: several works find no causal relation between the model's CoT and final answer~\citep{turpin2023languagemodelsdontsay,wang2025comprehensivesurveytrustworthinessreasoning}, casting doubt on the notion that the generated tokens allow the model to \textit{think} before answering. This behavior is less common, but still prominent in Large Reasoning Models (LRMs)~\citep{chua2025deepseekr1reasoningmodels}. Thus, long intermediate chains may not directly influence response quality~\citep{stechly2025beyond, kambhampati_stop_2025}, sparking interest in generating shorter intermediate tokens~\citep{arora2025traininglanguagemodelsreason,sui2025stopoverthinkingsurveyefficient}. We quantify the impact of deeper thinking on verifier quality through a controlled ablation study.

\paragraph{Setup.} We evaluate the impact of generating thinking traces on verifier quality by comparing short chain-of-thought (CoT) with longer reasoning-style traces. We also study the impact of varying \texttt{B$_{\text{tr}}\!=\!\length(\rvz)$}: the maximum permitted length of reasoning traces during training. We train four models: \texttt{GRPO-Instruct} with \texttt{B$_{\text{tr}}\!\,=\!\,$4096}, and three \texttt{GRPO-Think} variants with \texttt{B$_{\text{tr}}\!\,\in$\,\{4096, 8192, 16384\}}. \texttt{GRPO-Instruct} is initialized from the \texttt{Qwen2.5-Instruct}, which does not generate thinking traces by default. We report BoN and RL performance trends using the metrics discussed in \cref{sec: metrics}, along with the average costs per step, assuming \texttt{\$10.6} per H200-hour\footnote{\href{https://docs.jarvislabs.ai/blog/h200-price}{\scalerel*{\includegraphics{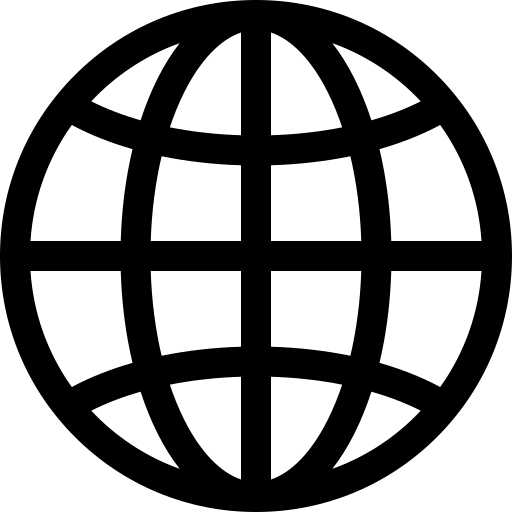}}{\textbf{C}} \texttt{jarvislabs.ai/h200-price}}}. We also study the effects of self-consistency~\citep{wang2023selfconsistencyimproveschainthought} on BoN performance.

\paragraph{Findings as a Best-of-N selector.} \texttt{GRPO-Think-16k} outperforms all other verifiers in \cref{tab: think-toggle-merged}, but the gap to \texttt{GRPO-Instruct} varies with scale. \texttt{GRPO-Instruct} and \texttt{GRPO-Think-4k} differ by $\leq$\,\texttt{2.4} BoN points on average for \texttt{1.5-7B} verifiers, indicating the \emph{style} of intermediate trace makes little difference at smaller scale. At \texttt{14B}, the same comparison goes to \texttt{8.4} points. Expanding the training reasoning budget follows the same pattern: for \texttt{1.5B}, the \texttt{4k}$\to$\texttt{8k} doubling adds \texttt{4.8} BoN points, but \texttt{8k}$\to$\texttt{16k} adds only \texttt{1.8}, indicating diminishing returns. In contrast, \texttt{7-14B} models keep climbing up to \texttt{16k}, with gains of \texttt{8.3} and \texttt{11.6} BoN points respectively at the \texttt{8k}$\to$\texttt{16k} step. These trends are likely driven by larger models being generally better at utilizing thinking primitives~\citep{DBLP:journals/corr/abs-2503-01307} and longer contexts~\citep{hsieh2024ruler, liu2024longgenbench}.

Across sizes, the verifiers stay robust to shifts in generator capabilities on \rqone{}, losing only $\approx$\,\texttt{5.2} BoN points across all models. This reproduces the scalable-supervision pattern of~\citet{burnsWeaktoStrongGeneralizationEliciting2023} and contrasts with earlier reports on verification tasks~\citep{DBLP:journals/corr/abs-2509-17995}. Similar to the average trend, the style of intermediate traces has minimal influence on \texttt{1.5-7B} models, and increasing \texttt{B$_\text{tr}$} to \texttt{8k} and \texttt{16k} tracks the same trend for shifts in generator capability.

Overall, the \texttt{1.5B} models are sensitive to adversarial perturbations, performing worse than the random baseline for \texttt{B$_\text{tr}$=4k}. This drop in performance is mitigated by scaling the reasoning budget to \texttt{16k}, increasing the model size, or both. Crucially, simply switching the \textit{style} of thinking traces has only a small impact on performance, and can even hurt the \texttt{7B} model. We observe robustness to adversarial perturbations only at the larger reasoning budgets and model sizes, contradicting the general notion that reasoning traces make models less vulnerable to judging biases~\citep{wangAssessingJudgingBias2025}. The training\,--\,evaluation partition in \workname{} allows us to attribute these trends directly to the training recipe, without contamination confounds.

We observe consistently large performance drops on \rqtwo{}, suggesting that easy-to-hard generalization performance is challenging for all model sizes. Contrary to the general trend, simply switching from CoT to reasoning traces improves BoN performance across the board, with additional scaling bringing linear gains. Since the codes in \rqtwo{} are highly similar (c.f. \cref{tab: dset-stats}), verifiers benefit from utilizing reasoning primitives like backtracking and subgoal-setting~\citep{DBLP:journals/corr/abs-2503-01307}.

\begin{figure}[!t]
    \begin{minipage}[c]{0.48\linewidth}
        \centering
        \includegraphics[width=\linewidth]{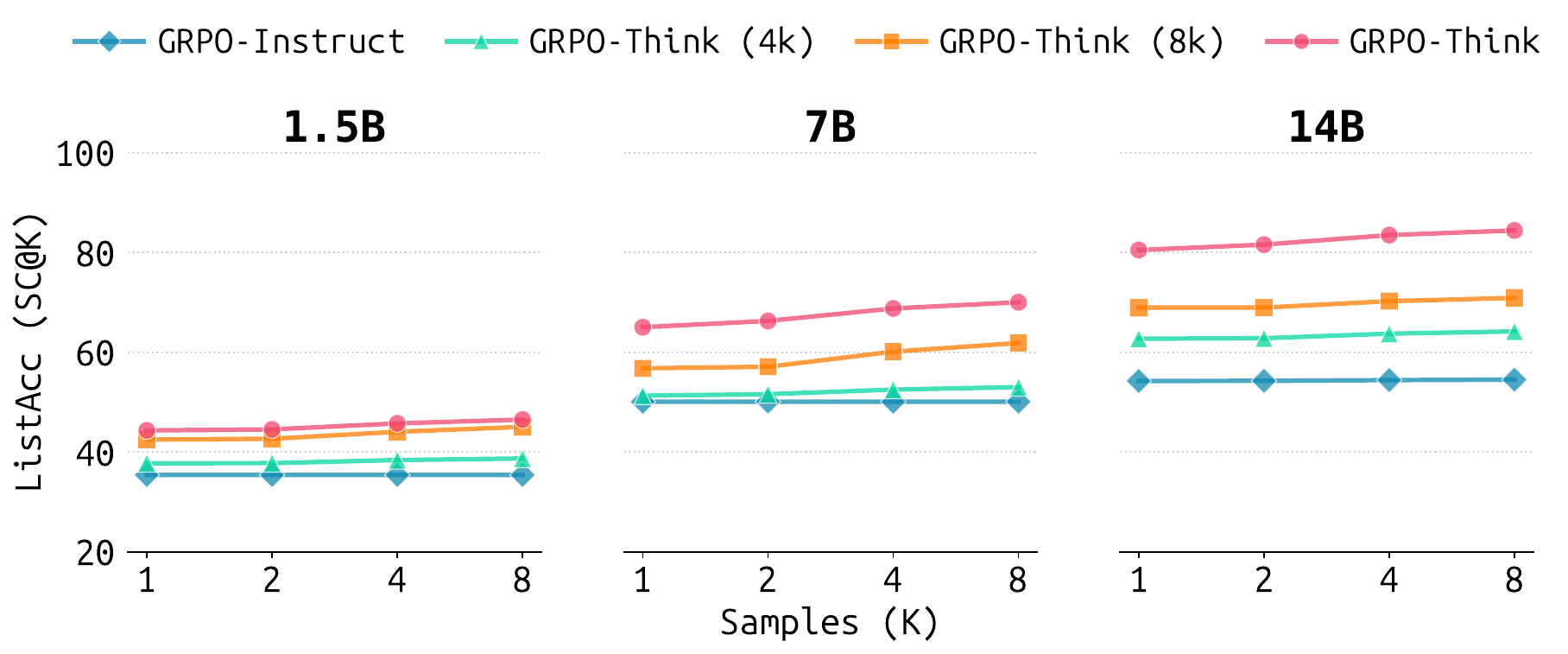}
        \caption{\textbf{Inference-time scaling for the thinking ablation.} CoT-trained models don't benefit from additional compute, while thinking models see a small upward trend across all sizes.}
        \label{fig: sc-curves-think}
    \end{minipage}
    \hfill
    \begin{minipage}[c]{0.48\linewidth}
        \centering
        \includegraphics[width=\linewidth]{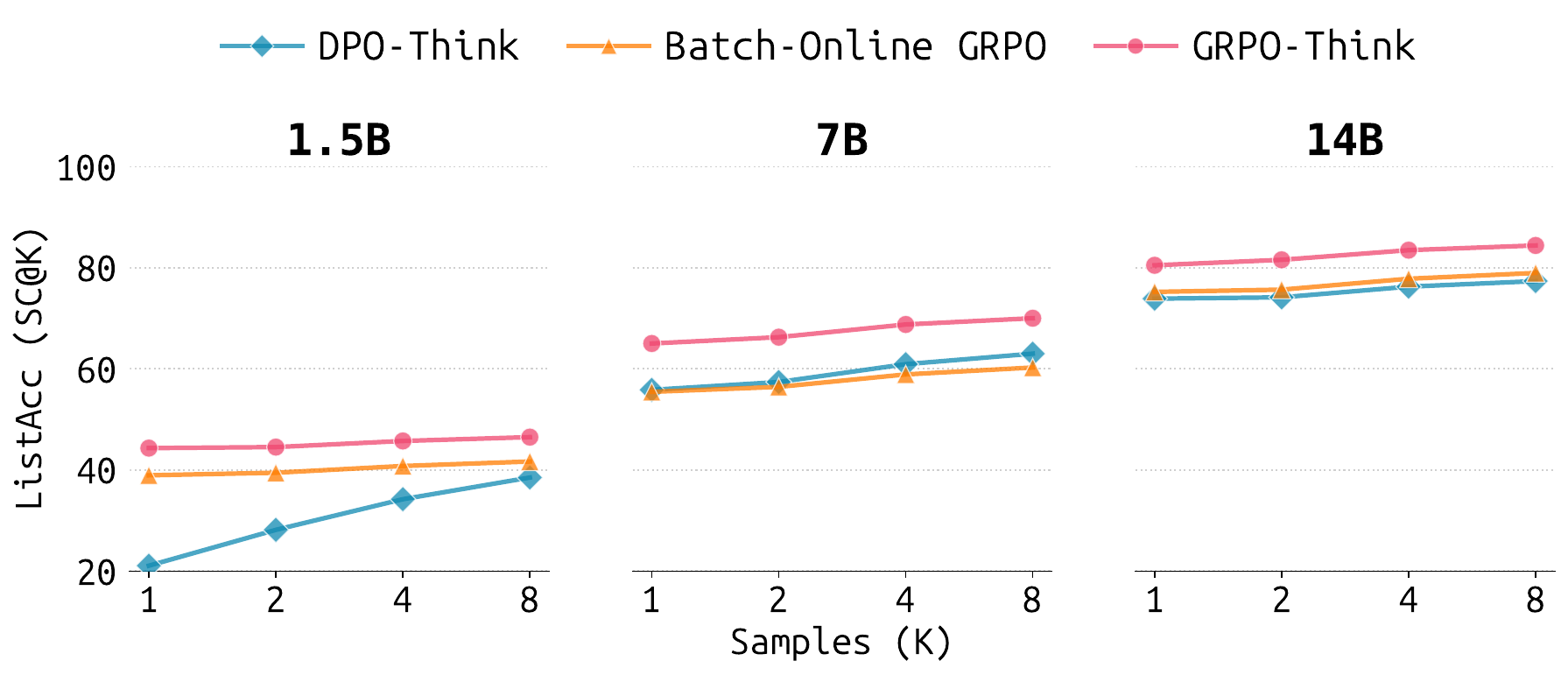}
        \caption{\textbf{Inference-time scaling for the online-offline ablation.} The offline\,--\,online gap narrows with increasing \texttt{K}, but does not close. \texttt{DPO-Think} can surpass \texttt{BO-GRPO} at larger inference budgets.}
        \label{fig: sc-curves-online}
    \end{minipage}
\end{figure}

Thinking is also vital for utilizing additional compute at test-time with self-consistency, with CoT traces in \texttt{GRPO-Instruct} yielding a flat \texttt{SC@K} curve (\cref{fig: sc-curves-think}). All \texttt{GRPO-Think} variants see a slight upward trend, with the effect being most pronounced at \texttt{B$_\text{tr}$=16k}. The curves are strictly ordered across all sizes, suggesting that while self-consistency provides modest gains at inference-time, it cannot replace \texttt{B$_\text{tr}$} scaling.

\paragraph{Findings as an RL reward.} The comparison under \ktau{} paints a very similar picture to the one under \texttt{ListAcc}, with \texttt{GRPO-Think-16k} emerging as the best verifier for RL training with \texttt{\ktau=52.03} at the \texttt{14B} scale. The training budget sweep also tells the same story under \ktau{} with the algorithm that produces the best selector also generally producing the best ranker, albeit with different magnitudes. Similarly, the trends established for the OOD evaluation datasets on BoN also generally hold for the RL setting. However, there are some notable exceptions to this trend. Although at \texttt{1.5B}, increasing \texttt{B$_\text{tr}$} from \texttt{8k}$\to$\texttt{16k} yields diminishing returns for BoN as established earlier, it yields continued gains for RL. On \rqfour{}, this scaling yields a modest \texttt{2.88} BoN points, but a more substantial \texttt{6.95} \ktau{} points. Surprisingly, while increasing \texttt{B$_\text{tr}$} from \texttt{4k}$\to$\texttt{8k} for the \texttt{7B} model brings a \texttt{5.49} BoN point gain on \heldout{}, it decreases \ktau{} by \texttt{2.18}.

\subsection{\texttt{RQ2}: Is On-policy Learning Essential for Verifier Training?}
\label{sec: online-toggle}
\begin{table*}[!t]
    \centering
    \caption{\textbf{Results for ablating on-policy learning (\raisebox{-0.2em}{\includegraphics[height=1em]{Figures/online.png}}).} We report list accuracy and Kendall $\tau$ scores for BoN and RL, respectively. Online learning is important for small verifiers, but its importance diminishes as scale increases. Batch-online methods are similarly useful at small scales but don't help larger models. On-policy is especially irrelevant for \easy{}-to-\hard{} generalization. Subscripts are 95\% confidence intervals.}
    \small
    \tt
    \setlength{\tabcolsep}{3pt}
    \scalebox{0.8}{
        \begin{tabular}{clcc|w{c}{1.3cm}w{c}{1.3cm}|w{c}{1.3cm}w{c}{1.3cm}|w{c}{1.3cm}w{c}{1.3cm}|w{c}{1.3cm}w{c}{1.3cm}|w{c}{1.1cm}w{c}{1.1cm}}
            \multicolumn{14}{c}{\thinking{}~\texttt{\textbf{Thinking}} \quad \negatives{}~\texttt{\textbf{Negatives}} \quad \online{}~\texttt{\textbf{Online}} \quad \semionline{}~\texttt{\textbf{Batch-online}}}                                                                                                                                                                                                                                                                              \\
            \toprule
                                                     & \multirow{2}{*}{\textbf{Method}} & \multirow{2}{*}{\textbf{Size}} & \multirow{2}{*}{\textbf{Cost (\$)}} & \multicolumn{2}{c|}{\textbf{\heldout{}}} & \multicolumn{2}{c|}{\textbf{\rqone{}}} & \multicolumn{2}{c|}{\textbf{\rqtwo{}}} & \multicolumn{2}{c|}{\textbf{\rqfour{}}} & \multicolumn{2}{c}{\textbf{Average}}                                                                                                                    \\
            \cmidrule(lr){5-6} \cmidrule(lr){7-8} \cmidrule(lr){9-10} \cmidrule(lr){11-12} \cmidrule(lr){13-14}
                                                     &                                  &                                &                                     & \texttt{BoN}                             & \texttt{RL}                            & \texttt{BoN}                           & \texttt{RL}                             & \texttt{BoN}                         & \texttt{RL}          &
            \texttt{BoN}                             & \texttt{RL}                      & \texttt{BoN}                   & \texttt{RL}                                                                                                                                                                                                                                                                                                                                                          \\ \midrule
            -                                        & Random                           &                                & -                                   & \scval{32.08}                            & \rcval{0.00}                           & \scval{32.08}                          & \rcval{0.00}                            & \scval{32.08}                        & \rcval{0.00}         & \scval{32.08}        & \rcval{0.00}         & \scval{32.08}        & \rcval{0.00}         \\ \midrule
            \thinking{}\,\negatives{}\,\noonline{}   & DPO-Think                        & \multirow{3}{*}{1.5B}          & ~5.951$^{*}$                        & \scvalc{21.70}{0.39}                     & \rcvalc{9.46}{2.63}                    & \scvalc{19.90}{0.35}                   & \rcvalc{16.00}{2.61}                    & \scvalc{23.41}{0.37}                 & \rcvalc{12.08}{2.63} & \scvalc{19.37}{0.15} & \rcvalc{16.39}{2.58} & \scvalc{21.11}{0.32} & \rcvalc{13.48}{2.61} \\
            \thinking{}\,\negatives{}\,\semionline{} & BO-GRPO                          &                                & 4.322                               & \scvalc{43.13}{0.55}                     & \rcvalc{17.66}{2.48}                   & \scvalc{39.89}{0.52}                   & \rcvalc{12.53}{2.45}                    & \scvalc{39.26}{0.54}                 & \rcvalc{7.19}{2.49}  & \scvalc{33.83}{0.22} & \rcvalc{6.48}{2.54}  & \scvalc{38.99}{0.46} & \rcvalc{10.97}{2.49} \\
            \thinking{}\,\negatives{}\,\online{}     & GRPO-Think                       &                                & 7.806                               & \scvalc{49.58}{0.65}                     & \rcvalc{27.49}{2.32}                   & \scvalc{46.09}{0.63}                   & \rcvalc{23.57}{2.33}                    & \scvalc{40.74}{0.64}                 & \rcvalc{8.92}{2.48}  & \scvalc{40.97}{0.26} & \rcvalc{16.45}{2.53} & \scvalc{44.38}{0.55} & \rcvalc{19.11}{2.42} \\ \midrule
            \thinking{}\,\negatives{}\,\noonline{}   & DPO-Think                        & \multirow{3}{*}{7B}            & ~6.403$^{*}$                        & \scvalc{63.75}{0.57}                     & \rcvalc{37.58}{2.05}                   & \scvalc{55.54}{0.57}                   & \rcvalc{31.89}{2.37}                    & \scvalc{51.20}{0.62}                 & \rcvalc{16.56}{2.40} & \scvalc{52.94}{0.25} & \rcvalc{28.81}{2.38} & \scvalc{55.88}{0.50} & \rcvalc{28.71}{2.30} \\
            \thinking{}\,\negatives{}\,\semionline{} & BO-GRPO                          &                                & 9.588                               & \scvalc{64.71}{0.60}                     & \rcvalc{41.27}{2.08}                   & \scvalc{56.18}{0.61}                   & \rcvalc{30.72}{2.31}                    & \scvalc{52.08}{0.66}                 & \rcvalc{15.68}{2.39} & \scvalc{49.35}{0.27} & \rcvalc{26.05}{2.44} & \scvalc{55.46}{0.54} & \rcvalc{28.43}{2.30} \\
            \thinking{}\,\negatives{}\,\online{}     & GRPO-Think                       &                                & 15.101                              & \scvalc{74.81}{0.57}                     & \rcvalc{51.52}{1.73}                   & \scvalc{67.28}{0.60}                   & \rcvalc{46.30}{2.03}                    & \scvalc{53.11}{0.69}                 & \rcvalc{20.31}{2.44} & \scvalc{65.04}{0.26} & \rcvalc{46.16}{2.14} & \scvalc{65.05}{0.53} & \rcvalc{41.07}{2.08} \\ \midrule
            \thinking{}\,\negatives{}\,\noonline{}   & DPO-Think                        & \multirow{3}{*}{14B}           & ~7.087$^{*}$                        & \scvalc{82.56}{0.52}                     & \rcvalc{51.12}{1.80}                   & \scvalc{74.39}{0.58}                   & \rcvalc{43.46}{1.85}                    & \scvalc{67.58}{0.68}                 & \rcvalc{29.90}{2.27} & \scvalc{71.06}{0.26} & \rcvalc{44.26}{2.10} & \scvalc{73.89}{0.51} & \rcvalc{42.19}{2.00} \\
            \thinking{}\,\negatives{}\,\semionline{} & BO-GRPO                          &                                & 31.144                              & \scvalc{83.82}{0.50}                     & \rcvalc{49.42}{1.75}                   & \scvalc{76.33}{0.56}                   & \rcvalc{41.31}{1.91}                    & \scvalc{67.34}{0.68}                 & \rcvalc{32.21}{2.22} & \scvalc{73.45}{0.25} & \rcvalc{42.01}{2.06} & \scvalc{75.29}{0.50} & \rcvalc{41.24}{1.98} \\
            \thinking{}\,\negatives{}\,\online{}     & GRPO-Think                       &                                & 36.992                              & \scvalc{88.02}{0.45}                     & \rcvalc{60.10}{1.64}                   & \scvalc{83.65}{0.49}                   & \rcvalc{60.21}{1.74}                    & \scvalc{66.84}{0.70}                 & \rcvalc{30.95}{2.26} & \scvalc{83.67}{0.21} & \rcvalc{56.84}{1.63} & \scvalc{80.54}{0.46} & \rcvalc{52.03}{1.82} \\ \bottomrule
        \end{tabular}
    }
    \label{tab: online-toggle-merged}
\end{table*}

\begin{finding_bon}
    \scriptsize
    \begin{itemize}[leftmargin=*, noitemsep]
        \item \texttt{Off-policy training collapses below random at 1.5B, but is competitive with semi-online methods at larger sizes.}
        \item \texttt{The online-offline gap between DPO--GRPO narrows with scale but never closes: from 23.27 BoN points at 1.5B to 6.65 at 14B.}
        \item \texttt{BO-GRPO recovers neither GRPO-Think's BoN performance nor offers a meaningful cost saving.}
        \item \texttt{Inference-time scaling narrows the offline--online BoN gap, most visibly at 1.5B, but cannot close it.}
    \end{itemize}
\end{finding_bon}
\begin{finding_rl}
    \scriptsize
    \begin{itemize}[leftmargin=*, noitemsep]
        \item \texttt{DPO-Think-1.5B's high \ktau{} scores are an artifact of noisy evaluation due to low response parseability.}
        \item \texttt{The necessity of on-policy learning for RL training similarly diminishes with scale.}
        \item \texttt{On-policy training has little effect on easy-to-hard generalization.}
        \item \texttt{BO-GRPO offers no consistent \ktau{} gain over offline DPO at 7\,--\,14B despite higher compute cost.}
    \end{itemize}
\end{finding_rl}

\paragraph{Background.} On-policy learning is perhaps the most widely studied and the most expensive aspect of RLVR training. Despite its effectiveness, on-policy training is very inefficient and often impractical. Thus, practitioners usually resort to introducing some amount of off-policyness to increase training efficiency~\citep{DBLP:conf/iclr/NoukhovitchHXHA25, DBLP:journals/corr/abs-2509-19128}. There is no consensus on its necessity: some works find it vital to success in RL algorithms~\citep{DBLP:conf/iclr/NoukhovitchHXHA25, tang2024understanding,yu2025optimizing}, while others claim that introducing a certain amount of off-policyness can match or even outperform fully on-policy methods on mathematical reasoning tasks~\citep{lanchantinBridgingOfflineOnline2025, chen2025retaining, song2024importance}.

\paragraph{Setup.} We study the impact of this decision through three representative algorithms. \texttt{DPO-Think} serves as our purely offline algorithm, and \texttt{Batch-online\,(BO-)\,GRPO} represents the middle ground between online and offline methods, sampling a batch of responses and performing multiple gradient updates on mini-batches of generated data~\citep{zhengGroupSequencePolicy2025}. We present the results in \cref{tab: online-toggle-merged}. For \texttt{DPO-Think}, we include the costs of creating an offline preference dataset as detailed in \cref{sec: detailed-hparams}.

\renewcommand{\thefootnote}{\fnsymbol{footnote}}\setcounter{footnote}{1}%
\footnotetext{DPO includes the cost for creating the offline dataset.}%
\renewcommand{\thefootnote}{\arabic{footnote}}

\paragraph{Findings as a Best-of-N selector.} \texttt{GRPO-Think} continues to dominate the ablated variants at all model sizes. The offline-online gap narrows with scale, with the \texttt{DPO-GRPO} gap shrinking from \texttt{23.27} BoN points at \texttt{1.5B} to just \texttt{6.65} at \texttt{14B}. \texttt{DPO-Think} collapses at smaller sizes, performing worse than the random baseline due to degeneration and unparseable verdicts. Specifically, \texttt{DPO-Think-1.5B} produces a parseable final answer in only {\tt 43\%} of cases on average (see \cref{tab: parse-rates} for more details). \texttt{BO-GRPO} is a good tradeoff at smaller scales, with lower costs than a fully offline approach, and performance within \texttt{6} points of the full \texttt{GRPO-Think} recipe.

However, the story flips at \texttt{7--14B} sizes. \texttt{DPO-Think} closes the gap with \texttt{BO-GRPO} on BoN while also decreasing costs, and is \texttt{4.4}$\times$ cheaper at \texttt{14B}. \texttt{BO-GRPO} fails to recover \texttt{GRPO-Think}'s BoN, and its lower cost is not worth the performance drop, since cheaper alternatives exist. Our observation contradicts that of~\citet{lanchantinBridgingOfflineOnline2025}, who report batch-online methods that match or outperform fully online training on math tasks. We attribute the discrepancy to their use of \texttt{Llama-3.1-8B-Instruct} (no thinking traces) and the original GRPO recipe~\citep{shaoDeepSeekMathPushingLimits2024}, while we incorporate improvements detailed in \cref{sec: detailed-hparams}.

The \rqone{} and \rqfour{} evaluations track the general trend almost exactly, except for \texttt{DPO-Think-7B} being slightly more robust to adversarial prompts than \texttt{BO-GRPO}. On-policy training has no perceptible impact on easy-to-hard generalization at \texttt{7-14B}, suggesting that a well-curated offline preference dataset may be sufficient for such generalization. Although \texttt{DPO-Think-1.5B} gains \texttt{1.71} BoN points on \rqtwo{}, it is still worse than random chance and we thus do not consider it a meaningful gain.

Scaling inference-time compute benefits all methods and model sizes (\cref{fig: sc-curves-online}). Such scaling particularly benefits \texttt{DPO-Think-1.5B}, which almost matches \texttt{BO-GRPO}'s BoN at \texttt{K=8}. This behavior is largely driven by a higher parse rate due to increased sampling at inference time. At \texttt{7B}, additional inference-time compute even allows \texttt{DPO-Think} to surpass \texttt{BO-GRPO}'s BoN scores at \texttt{K=8}, but still trails \texttt{GRPO-Think} even at \texttt{K=1}. We thus conclude that, similar to Thinking, inference-time compute scaling cannot fully compensate for the lack of on-policy training, but can significantly narrow the gap, especially for semi-online methods.

\paragraph{Findings as an RL reward model.} Similar to BoN, on-policy training makes \texttt{GRPO-Think} the best reranker for all model sizes. Despite sub-random BoN performance, \texttt{DPO-Think-1.5B} has competitive \ktau{} values at its scale, even scoring highest on \rqtwo{} with \texttt{\ktau=12.08}. However, this number does not indicate that \texttt{DPO-Think-1.5B} is a capable reranker. Rather, it is a result of the two metrics handling unparseable verdicts differently. Under list accuracy for BoN, unparseable instances are considered incorrect, causing the method to perform worse than random chance. However, \ktau{} discards such instances, effectively treating them as ties and penalizing both candidates equally in a pairwise tournament. Consequently, \ktau{} is calculated over a smaller set of comparisons, and the resulting value is noisy. We report the parse rates for all algorithms in \cref{tab: parse-rates}, which clearly illustrates this behavior. This degeneration is specific to the \texttt{1.5B} verifier, possibly due to the model's limited capacity. The parse rates for \texttt{DPO-Think} recover at larger scales, suggesting that their competitive \ktau{} relative to \texttt{BO-GRPO} indicates genuine reranking competence.

\texttt{BO-GRPO} continues to be a poor substitute for \texttt{GRPO-Think} even as an RL reward function. Although \texttt{DPO-Think} slightly underperforms \texttt{BO-GRPO}'s BoN on average at \texttt{14B}, this trend flips on \ktau{}, further underscoring the shortcomings of semi-online training as a substitute for on-policy training. Unlike BoN, where on-policy learning has little impact on easy-to-hard generalization, the same cannot be said for the RL setting where fully offline methods lag behind the online and semi-online methods by up to \texttt{3.75\%}. The behaviors of the evaluated recipes on \rqone{} and \rqfour{} largely track the same trends as BoN on \ktau{}.

\begin{table*}[!t]
    \centering
    \caption{\textbf{Results for ablating learning from negatives (\raisebox{-0.2em}{\includegraphics[height=1em]{Figures/negatives.png}}).} We report list accuracy and Kendall $\tau$ scores for BoN and RL respectively. Negatives are consistently beneficial across all model sizes for BoN, but more important for large models during RL. \easy{}-to-\hard{} generalization is largely unaffected by negatives. Subscripts are 95\% confidence intervals.}
    \small
    \tt
    \setlength{\tabcolsep}{3pt}
    \scalebox{0.8}{
        \begin{tabular}{llcc|w{c}{1.3cm}w{c}{1.3cm}|w{c}{1.3cm}w{c}{1.3cm}|w{c}{1.3cm}w{c}{1.3cm}|w{c}{1.3cm}w{c}{1.3cm}|w{c}{1.1cm}w{c}{1.1cm}}
            \multicolumn{14}{c}{\thinking{}~\texttt{\textbf{Thinking}} \quad \negatives{}~\texttt{\textbf{Negatives}} \quad \online{}~\texttt{\textbf{Online}}}                                                                                                                                                                                                                                                                                                                                                 \\
            \toprule
                                                   & \multirow{2}{*}{\textbf{Method}} & \multirow{2}{*}{\textbf{Size}} & \multirow{2}{*}{\textbf{Cost (\$)}} & \multicolumn{2}{c|}{\textbf{\heldout{}}} & \multicolumn{2}{c|}{\textbf{\rqone{}}} & \multicolumn{2}{c|}{\textbf{\rqtwo{}}} & \multicolumn{2}{c|}{\textbf{\rqfour{}}} & \multicolumn{2}{c}{\makebox[1em][c]{\textbf{Average}}}                                                                                                                    \\
            \cmidrule(lr){5-6} \cmidrule(lr){7-8} \cmidrule(lr){9-10} \cmidrule(lr){11-12} \cmidrule(lr){13-14}
                                                   &                                  &                                &                                     & \texttt{BoN}                             & \texttt{RL}                            & \texttt{BoN}                           & \texttt{RL}                             & \texttt{BoN}                                           & \texttt{RL}          & \texttt{BoN}         & \texttt{RL}          & \texttt{BoN}         & \texttt{RL}          \\ \midrule
            -                                      & Random                           &                                & -                                   & \scval{32.08}                            & \rcval{0.00}                           & \scval{32.08}                          & \rcval{0.00}                            & \scval{32.08}                                          & \rcval{0.00}         & \scval{32.08}        & \rcval{0.00}         & \scval{32.08}        & \rcval{0.00}         \\ \midrule
            \thinking{}\,\nonegatives{}\,\online{} & RAFT                             & \multirow{2}{*}{1.5B}          & 4.167                               & \scvalc{34.76}{0.48}                     & \rcvalc{14.20}{2.38}                   & \scvalc{31.88}{0.44}                   & \rcvalc{12.31}{2.48}                    & \scvalc{33.67}{0.46}                                   & \rcvalc{9.06}{2.43}  & \scvalc{29.12}{0.19} & \rcvalc{12.25}{2.58} & \scvalc{32.30}{0.39} & \rcvalc{11.96}{2.47} \\
            \thinking{}\,\negatives{}\,\online{}   & GRPO-Think                       &                                & 7.806                               & \scvalc{49.58}{0.65}                     & \rcvalc{27.49}{2.32}                   & \scvalc{46.09}{0.63}                   & \rcvalc{23.57}{2.33}                    & \scvalc{40.74}{0.64}                                   & \rcvalc{8.92}{2.48}  & \scvalc{40.97}{0.26} & \rcvalc{16.45}{2.53} & \scvalc{44.38}{0.55} & \rcvalc{19.11}{2.42} \\ \midrule
            \thinking{}\,\nonegatives{}\,\online{} & RAFT                             & \multirow{2}{*}{7B}            & 6.948                               & \scvalc{60.86}{0.59}                     & \rcvalc{38.82}{2.11}                   & \scvalc{52.00}{0.58}                   & \rcvalc{31.89}{2.33}                    & \scvalc{48.84}{0.63}                                   & \rcvalc{17.33}{2.41} & \scvalc{49.24}{0.26} & \rcvalc{29.36}{2.48} & \scvalc{52.72}{0.52} & \rcvalc{29.35}{2.33} \\
            \thinking{}\,\negatives{}\,\online{}   & GRPO-Think                       &                                & 15.101                              & \scvalc{74.81}{0.57}                     & \rcvalc{51.52}{1.73}                   & \scvalc{67.28}{0.60}                   & \rcvalc{46.30}{2.03}                    & \scvalc{53.11}{0.69}                                   & \rcvalc{20.31}{2.44} & \scvalc{65.04}{0.26} & \rcvalc{46.16}{2.14} & \scvalc{65.05}{0.53} & \rcvalc{41.07}{2.08} \\ \midrule
            \thinking{}\,\nonegatives{}\,\online{} & RAFT                             & \multirow{2}{*}{14B}           & 12.906                              & \scvalc{75.55}{0.56}                     & \rcvalc{49.42}{1.82}                   & \scvalc{66.02}{0.61}                   & \rcvalc{41.31}{2.07}                    & \scvalc{65.23}{0.65}                                   & \rcvalc{32.21}{2.26} & \scvalc{62.03}{0.27} & \rcvalc{42.01}{2.11} & \scvalc{67.20}{0.52} & \rcvalc{41.24}{2.07} \\
            \thinking{}\,\negatives{}\,\online{}   & GRPO-Think                       &                                & 36.992                              & \scvalc{88.02}{0.45}                     & \rcvalc{60.10}{1.64}                   & \scvalc{83.65}{0.49}                   & \rcvalc{60.21}{1.74}                    & \scvalc{66.84}{0.70}                                   & \rcvalc{30.95}{2.26} & \scvalc{83.67}{0.21} & \rcvalc{56.84}{1.63} & \scvalc{80.54}{0.46} & \rcvalc{52.03}{1.82} \\ \bottomrule
        \end{tabular}
    }
    \label{tab: neg-toggle-merged}
\end{table*}

\subsection{\texttt{RQ3}: Do Negatives Benefit Code Verifiers?}
\label{sec: neg-toggle}

\begin{finding_bon}
    \scriptsize
    \begin{itemize}[leftmargin=*, noitemsep]
        \item \texttt{GRPO-Think outperforms RAFT at every scale with a near-constant gap of $\approx$12.6 BoN points.}
        \item \texttt{RAFT-1.5B barely exceeds the random baseline (32.30 vs. 32.08) and falls below it on \rqone{} and \rqfour{}.}
        \item \texttt{Inference-time scaling cannot substitute for negatives: RAFT at K=8 fails to match GRPO-Think even at K=1.}
    \end{itemize}
\end{finding_bon}
\begin{finding_rl}
    \scriptsize
    \begin{itemize}[leftmargin=*, noitemsep]
        \item \texttt{Unlike BoN, the \ktau{} gap grows from 7.2 at 1.5B to 10.8 at 14B, indicating a growing importance of negatives for RL performance.}
        \item \texttt{RAFT becomes increasingly unstable at larger sizes, with performance even degrading over training.}
        \item \texttt{Negatives provide no discriminative benefit on \rqtwo{}, with RAFT performing comparably to GRPO-Think.}
    \end{itemize}
\end{finding_rl}

\paragraph{Background.} Learning from negative samples is a characteristic of RL algorithms, as well as of contrastive methods like DPO~\citep{rafailovDirectPreferenceOptimization2024}, which optimize the RL objective directly. DPO suffers from reward over-optimization~\citep{gaoScalingLawsReward2023}, and~\citet{xuDPOSuperiorPPO2024} find that even iterative DPO fails to beat the SFT baseline. The literature on negatives is mixed:~\citet{arnal2025asymmetric} find that successes are more important than failures in an offline setup, whereas~\citet {zhu2025surprising} find negative reinforcement much more important, even outperforming full training in some scenarios.~\citet{xiongMinimalistApproachLLM2025} find that learning only from positives comes with a minor performance drop, and certain negative signals can be detrimental.

\begin{figure}[!t]
    \centering
    \begin{minipage}[c]{0.48\linewidth}
        \includegraphics[width=\linewidth]{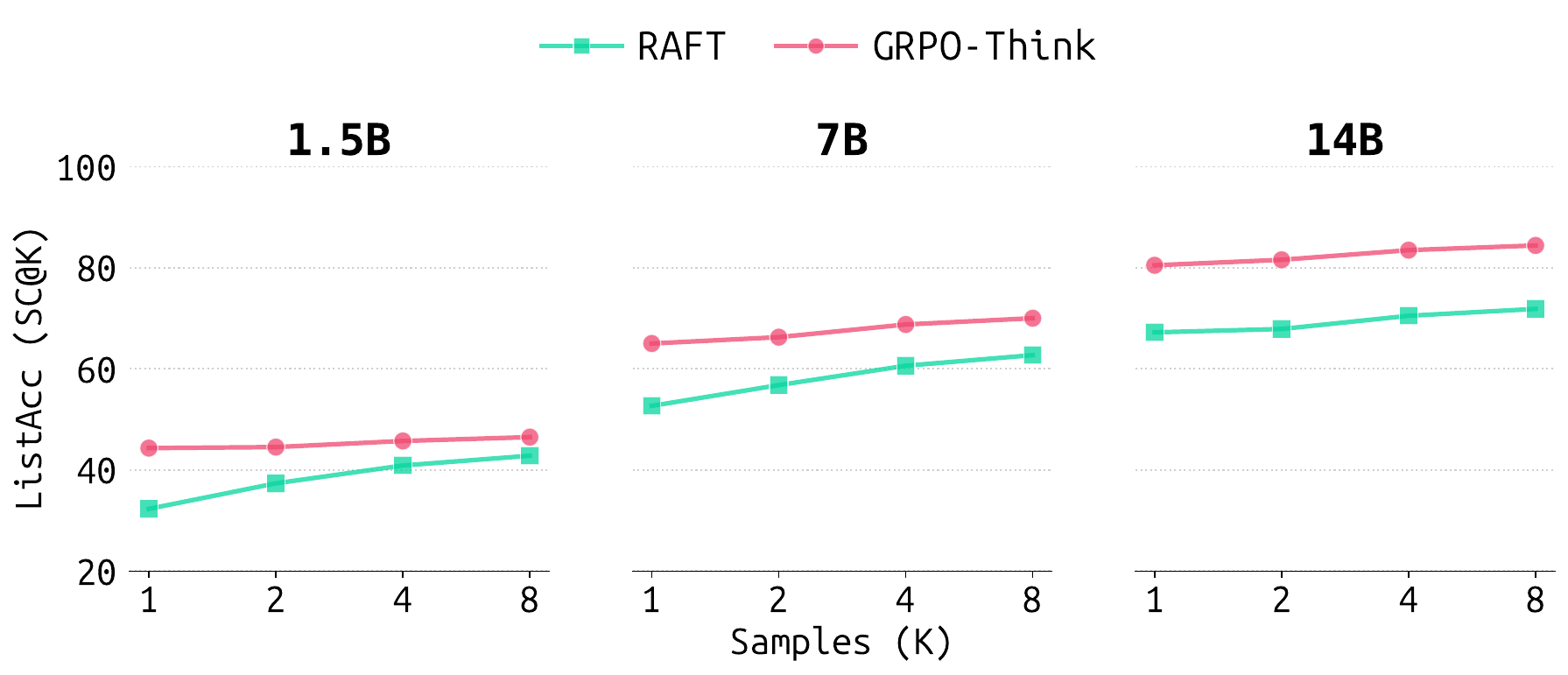}
        \caption{\textbf{Inference-time scaling plots for ablating negative samples.} The \texttt{RAFT-GRPO} gap narrows with compute, but is persistent across scales.}
        \label{fig: sc-curves-negatives}
    \end{minipage}
    \hfill
    \begin{minipage}[c]{0.48\linewidth}
        \includegraphics[width=\linewidth]{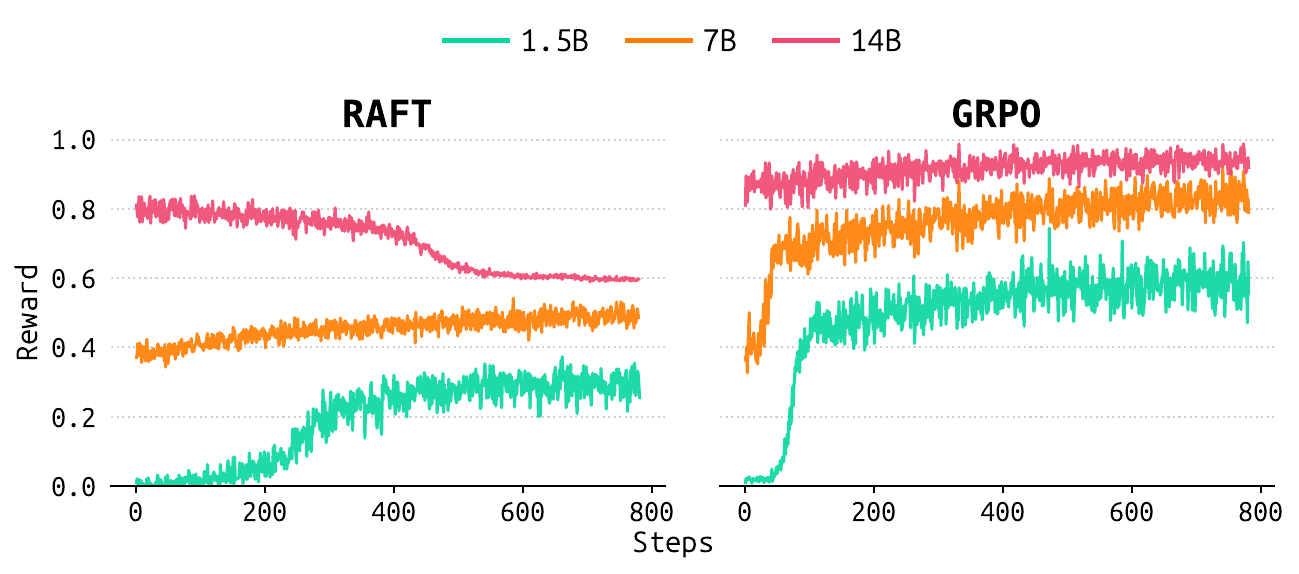}
        \caption{\textbf{Reward curves over training for \texttt{RAFT} and \texttt{GRPO}, respectively.} Training without negative samples is increasingly unstable for larger models.}
        \label{fig: raft-v-grpo-curve}
    \end{minipage}
\end{figure}

\paragraph{Setup.} We compare \texttt{GRPO} to a variant of \texttt{RAFT}~\citep{dongRAFTRewardRAnked}, modified to use verifiable rewards. \texttt{RAFT} samples and scores $N$ generations online, training on only the correct responses via next-token prediction.

\paragraph{Findings as a Best-of-N selector.} \texttt{GRPO-Think} outperforms \texttt{RAFT} across all model sizes on BoN with a near constant gap of \texttt{$\approx$\,12.6} points (\cref{tab: neg-toggle-merged}). This contrasts with the Online and Thinking components, whose importance diminishes and increases with scale, respectively. However, the gap to \texttt{GRPO-Think} is more consequential at smaller scales, with \texttt{RAFT-1.5B} scoring just above the random baseline of \texttt{32.08} points on average and dipping below it on \rqone{} and \rqfour{}.

The trends on individual evaluations track the average trend, with \texttt{GRPO-Think} consistently outperforming \texttt{RAFT} on BoN. Also, increasing inference compute cannot replace negative samples during training (\cref{fig: sc-curves-negatives}). While it narrows the gap to \texttt{GRPO-Think}, especially at \texttt{1.5B}, \texttt{RAFT} at \texttt{K=8} is worse than \texttt{GRPO-Think} at \texttt{K=1}.

\paragraph{Findings as an RL reward model.} While \texttt{GRPO-Think} remains the best reranker across model sizes, the \ktau{} evaluation tells a more nuanced story than BoN. The \texttt{RAFT-GRPO} gap grows from \texttt{7.2} \ktau{} points at \texttt{1.5B} to \texttt{10.8} at \texttt{14B}, suggesting a monotonic growth in the role of negative samples. The reason for this behavior is revealed by the training reward curves (\cref{fig: raft-v-grpo-curve}). Despite a steady growth at \texttt{1.5B}, the reward curves of \texttt{RAFT} flatline and even degrade at larger sizes, while \texttt{GRPO-Think} continues to improve. We further validate this behavior across other variants of \texttt{RAFT} in \cref{sec: extended-raft}.

\texttt{GRPO-Think} is more robust to shifts in generator capability and adversarial perturbations across all sizes. However, similar to our on-policy ablation, both \texttt{RAFT} and \texttt{GRPO-Think} rank near-correct distractors poorly and have largely comparable \ktau{} on \rqtwo{}.

\section{Optimality Analysis}
\label{sec: analysis}
In the previous section, we studied the individual roles of three components of RLVR training: Thinking, Negatives, and Online. Through our ablations, we established that all three components make nontrivial contributions to the overall success of \texttt{GRPO-Think}, which is the best-performing verifier on average in both BoN and RL applications. However, performance is not the only factor to consider while training verifiers, as each component has a disproportionate cost as seen in \cref{tab: think-toggle-merged,tab: online-toggle-merged,tab: neg-toggle-merged}.

\begin{figure}[h]
    \centering
    \includegraphics[width=\linewidth]{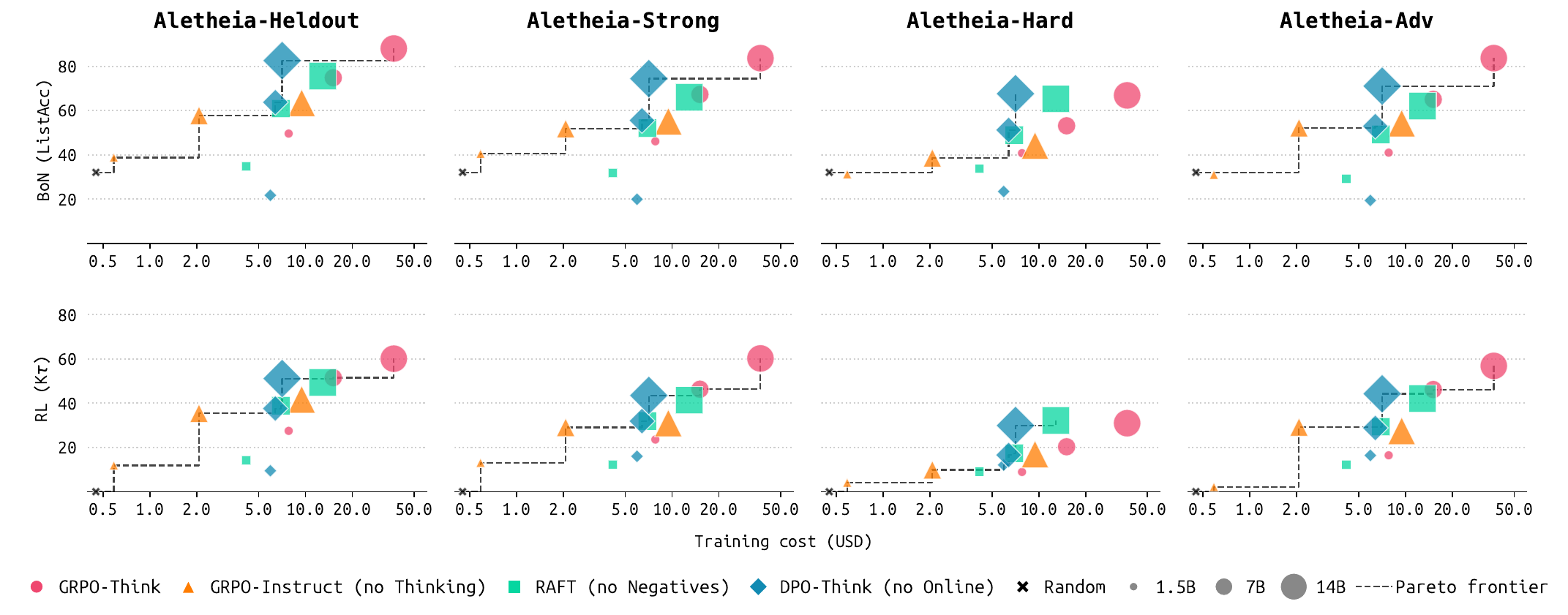}
    \caption{\textbf{Cost-performance Pareto curves for \texttt{BoN} and \texttt{RL} applications.} Top row: BoN (\texttt{ListAcc}); bottom row: RL (\ktau{}). Each column represents a different evaluation dataset. \texttt{DPO-Think-14B} is on the frontier in every panel, while \texttt{GRPO-Think-16k} is dominated on \rqtwo{} for all model sizes.}
    \label{fig: pareto}
\end{figure}

In this section, we study the trade-off between performance and cost for each of these axes. Specifically, we plot the full \texttt{GRPO-Think} recipe alongside the three ablated variants: \texttt{GRPO-Instruct} (No Thinking), \texttt{RAFT} (No Negatives), and \texttt{DPO-Think} (No Online) against training cost per step for the \texttt{BoN} and \texttt{RL} usecases on each \workname{} evaluation. The dashed line in \cref{fig: pareto} marks the empirical Pareto frontier per panel.

\noindent\textbf{Offline training achieves near-peak performance at one-fifth the cost of fully online training.}
\texttt{DPO-Think-14B} occupies a unique position across all eight panels: at \texttt{\$7.09} per step, it achieves 14B-scale performance at a \texttt{5.2$\times$} lower cost than \texttt{GRPO-Think-14B}, placing it on the Pareto frontier across all evaluations. \texttt{DPO-Think-1.5B} is dominated by other models due to the fixed cost of creating the offline preference dataset and poor performance stemming from a low parse rate (\cref{tab: parse-rates}).

\noindent\textbf{The \texttt{GRPO-Instruct} verifiers anchor the low-budget end, while \texttt{RAFT} is dominated on all BoN settings.}
The \texttt{1.5B} verifier is the cheapest method we study, but it is dominated by a random baseline in some BoN settings. \texttt{GRPO-Instruct-7B} has a good cost\,--\,performance tradeoff, and is on the Pareto frontier in all panels. In contrast, removing negatives yields neither the cost savings of \texttt{GRPO-Instruct} nor the performance of \texttt{DPO-Think}, making it fall below the frontier.

\noindent\textbf{\easy{}-to-\hard{} generalization favors cheaper alternatives over full GRPO-Think.}
\texttt{GRPO-Think} earns its high cost by extending the BoN and RL frontiers in-distribution and in evaluations with stronger generators and adversarial perturbations. However, \rqtwo{} is a notable exception, where \texttt{GRPO-Think} is dominated in both metrics. We conclude that thinking traces are the most vital for \easy{}-to-\hard{} generalization, as evidenced by \texttt{DPO-Think} and \texttt{RAFT} achieving (near-)Pareto-optimal performance on both metrics.

\noindent\textbf{BoN and RL frontiers largely agree on structure but have some important differences.}
Although the two metrics agree on the coarse frontier topology, they differ in three key ways. First, \texttt{GRPO-Think} enters the Pareto frontier on \heldout{} and \rqfour{} for RL by narrowly outperforming \texttt{DPO-Think-14B}, but is dominated on BoN. This suggests that on-policy training provides a larger benefit for pairwise \textit{ranking} than for argmax \textit{selection}. Second, the magnitude of the \texttt{GRPO-Think-14B} premium differs by metric: on \heldout{}, it extends the frontier by \texttt{5.5 ListAcc} points but by \texttt{9.0} \ktau{} points beyond \texttt{DPO-Think-14B}, suggesting pairwise ranking quality benefits disproportionately from on-policy training compared to argmax selection. Lastly, \texttt{RAFT-14B} is clearly dominated in all BoN evaluations but is near-optimal on RL, even entering the Pareto frontier on \rqtwo{}, which suggests that negative samples are disproportionately effective for improving ranking performance over argmax selection.


\section{Downstream Integration}
\label{sec: downstream}
In the previous sections, we used the \workname{} testbed to analyze different components of the RLVR recipe for verifier training, and established several scale-dependent takeaways and recommendations for practitioners. In this section, we evaluate the integrity of these findings by testing our verifiers in a setting more aligned with real-world applications: Best-of-N selection. Specifically, we generate \texttt{16} solutions each with temperature \texttt{0.8} using \texttt{Qwen2.5-Coder-7B-Instruct}~\citep{qwen2.5} on \texttt{1055} LiveCodeBench problems published between May 2023 and April 2025~\citep{DBLP:conf/iclr/JainHGLYZWSSS25}, and deploy all \texttt{21} studied verifiers to select the best solution. Using a mid-range generator at high temperature allows us to evaluate the performance in a non-trivial setting where the 16 generated solutions are of varying quality. We additionally compare the performance of our verifiers on two external reward model benchmarks: RM-Bench~\citep{liuRMBenchBenchmarkingReward2024} and Themis-CodeRewardBench (CRB; \citealp{DBLP:journals/corr/abs-2605-00754}). We then measure the correlation between the oracle verifier ranking at each scale as obtained via BoN\footnote{Global rank correlation is inflated because all proxies rank larger models better, and doesn't reflect algorithmic differences.}, and the ranking produced by the various proxies in \cref{tab: downstream_correlation} (Left) (exact performance numbers are in \cref{tab: downstream_performance}).

\begin{table}[!t]
    \centering
    \caption{\textbf{Correlation between Oracle (BoN) and Proxy Rankings. (Left)} We report the tie-corrected Kendall's $\tau$-b correlation and Spearman's $\rho$ between the ranking of verifiers based on their BoN performance and the ranking based on various proxy metrics. An Aletheia metric always dominates, but the results are unreliable due to noisy BoN rankings. \textbf{(Right)} We report the percentage of distinguishable comparisons correctly identified by each proxy metric at $\alpha\!\,=\!\,0.05$ and $\alpha\!\,=\!\,0.01$ using the paired McNemar test.}
    \label{tab: downstream_correlation}
    \tt
    \small
    \scalebox{0.8}{
        \begin{tabular}{clccc||cc}
            \toprule
                                                                                         & \textbf{Proxy}               & \textbf{1.5B $\tau$-b ($\rho$)} & \textbf{7B $\tau$-b ($\rho$)} & \textbf{14B $\tau$-b ($\rho$)} & \raisebox{0.2\height}{$\alpha$}\,=\,0.05 (n=37) & \raisebox{0.2\height}{$\alpha$}\,=\,0.01 (n=25) \\\midrule
            \raisebox{-0.25\height}{\includegraphics[height=1.5em]{Figures/greek2.png}}  & Aletheia ListAcc (Avg)       & +0.52 (+0.68)                   & \textbf{+0.90 (+0.96)}        & +0.88 (+0.95)                  & 94.6\%                                          & \textbf{100\%}                                  \\
            \raisebox{-0.25\height}{\includegraphics[height=1.5em]{Figures/greek2.png}}  & Aletheia \ktau{} (Avg)       & \textbf{+0.81 (+0.89)}          & +0.62 (+0.75)                 & \textbf{+0.98 (+0.99)}         & \textbf{100\%}                                  & \textbf{100\%}                                  \\
            \raisebox{-0.25\height}{\includegraphics[height=1.5em]{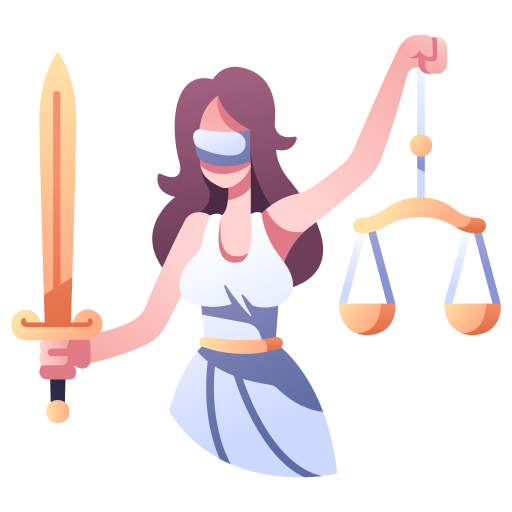}}  & CRB (Overall)                & +0.14 (+0.36)                   & +0.43 (+0.71)                 & +0.59 (+0.72)                  & 78.4\%                                          & 80\%                                            \\
            \raisebox{-0.25\height}{\includegraphics[height=1.5em]{Figures/themis.png}}  & CRB (Functional Correctness) & -0.14 (-0.21)                   & +0.24 (+0.39)                 & \textbf{+0.98 (+0.99)}         & 70.3\%                                          & 68\%                                            \\
            \raisebox{-0.25\height}{\includegraphics[height=1.3em]{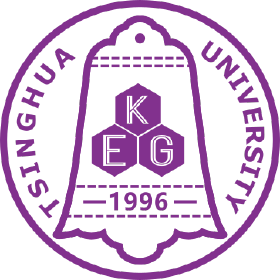}} & RMBench (Overall)            & +0.43 (+0.64)                   & +0.71 (+0.86)                 & \textbf{+0.98 (+0.99)}         & 91.9\%                                          & 96\%                                            \\
            \raisebox{-0.25\height}{\includegraphics[height=1.3em]{Figures/rmbench.png}} & RMBench (Code)               & +0.39 (+0.45)                   & +0.62 (+0.75)                 & +0.88 (+0.95)                  & 89.2\%                                          & 96\%                                            \\\bottomrule
        \end{tabular}
    }
\end{table}

Clearly, the best proxy at all scales is one of our proposed metrics, with \texttt{ListAcc} being the best at \texttt{7B} and \ktau{} being the best at \texttt{1.5B} and \texttt{14B}. Overall, the \texttt{14B} verifiers seem the easiest to separate, while the \texttt{1.5B} verifiers are the hardest. However, this analysis suffers an important caveat: the BoN rankings are compressed in a small range of \texttt{[16.59, 24.83]}, with random chance scoring \texttt{15.63} and oracle sandboxed execution reaching \texttt{26.16} several comparisons are within noise of each other. While a potential fix is to rerun the BoN evaluation multiple times to gain a more reliable estimate of the true BoN ranking, this is impractical to do at a meaningful scale. Instead, we use the paired McNemar test~\citep{mcnemar1947note} to distinguish between verifiers that are statistically significantly different from each other. Each verifier chooses among the same candidates for each problem, and McNemar's test on the discordant pairs tells us which pair of verifiers can truly be distinguished from each other based on downstream performance. We then evaluate each proxy on its ability to predict every ordering the deployment statistically establishes in \cref{tab: downstream_correlation} (Right). Both Aletheia metrics outperform existing benchmarks in distinguishing between statistically significant differences, with \ktau{} achieving perfect performance at both $\alpha$ levels. Additionally, we see that our findings from \cref{sec: results} drawn from \workname{} are fully supported by downstream BoN performance as well as by RM-Bench. Only Finding F3 is not supported by CRB's ranking, which is likely due to its construction spanning several auxillary tasks of code generation and non-competition code, which our verifiers were not trained on. Overall, these results suggest that our findings are valid and generalize to real-world downstream performance, and that our proposed metrics are the best proxies for evaluating verifier performance.

\begin{table}[!t]
    \centering
    \caption{\textbf{Validity of Aletheia's findings across BoN and external benchmarks.} We study the validity of the findings from \cref{sec: results} by checking if they hold across BoN, CRB, and RM-Bench. We report the rule used to validate each finding, and whether it holds for each benchmark. Overall, our findings are well-supported by oracle BoN rankings and external benchmarks.}
    \label{tab: findings_validity}
    \tt
    \small
    \scalebox{0.8}{
        \begin{tabular}{rlcccc}
            \toprule
                        & \textbf{Claim}                                                 & \textbf{Rule}                                                                                                                                                                      & \textbf{BoN} & \raisebox{-0.25\height}{\includegraphics[height=1.5em]{Figures/themis.png}} & \raisebox{-0.25\height}{\includegraphics[height=1.5em]{Figures/rmbench.png}} \\\midrule
            \vspace{0.4em}
            \textbf{F1} & The importance of thinking traces grows with scale             & $\Delta^{\nothinking{}}_{\thinking{}}$(1.5B)\,<\,$\Delta^{\nothinking{}}_{\thinking{}}$(7B)\,<\,$\Delta^{\nothinking{}}_{\thinking{}}$(14B)                                        & \tabletick{} & \tabletick{}                                                                & \tabletick{}                                                                 \\\vspace{0.4em}
            \textbf{F2} & The importance of on-policy learning decreases with scale      & $\Delta^{\noonline{}}_{\online{}}$(1.5B)\,>\,$\Delta^{\noonline{}}_{\online{}}$(7B)\,>\,$\Delta^{\noonline{}}_{\online{}}$(14B)                                                    & \tabletick{} & \tabletick{}                                                                & \tabletick{}                                                                 \\\vspace{0.4em}
            \textbf{F3} & Scaling reasoning budget beyond 8k benefits larger models more & $\Delta^{8\text{k}}_{16\text{k}}$(1.5B)\,<\,$\Delta^{8\text{k}}_{16\text{k}}$(7B)\,<\,$\Delta^{8\text{k}}_{16\text{k}}$(14B)                                                       & \tabletick{} & \tablecross{}                                                               & \tabletick{}                                                                 \\
            \textbf{F4} & Batch-online has no significant gain over offline at 7-14B     & $\Delta^{\noonline{}}_{\semionline{}}$(1.5B)\,>\,$\epsilon$; $\Delta^{\noonline{}}_{\semionline{}}$(7B)\,<\,$\epsilon$; $\Delta^{\noonline{}}_{\semionline{}}$(14B)\,<\,$\epsilon$ & \tabletick{} & \tabletick{}                                                                & \tabletick{}                                                                 \\
            \bottomrule
        \end{tabular}
    }
\end{table}

\section{Related Work}
\label{sec: related}
We briefly elaborate on the three most relevant lines of existing work: (1) RLVR for verifier models, (2) surrogate code execution verifiers, and (3) prior analyses of RL components in LLMs.

\paragraph{RLVR for LLM verifiers.}
Recent literature has substantially expanded verifier training by framing reward modeling as a verifiable re-ranking reasoning task~\citep{DBLP:journals/corr/abs-2505-10320,DBLP:journals/corr/abs-2504-00050,DBLP:journals/corr/abs-2505-14268}. Such models have demonstrated state-of-the-art performance on popular reward model benchmarks and have been integrated into the production post-training pipelines of several modern LLMs~\citep{DBLP:journals/corr/abs-2504-13914, team2025kimi, nvidia_nemotron_nano_v3_2025}. Despite empirical gains, the optimal configuration for training such models remains under-explored. In this work, we uncover a practical cost-performance guide for verifier training across three disparate model sizes by ablating three core components of the RLVR recipe.

\paragraph{Surrogate code execution.}
LLMs as surrogate code executors have taken several forms, including regression-based scoring models~\citep{DBLP:conf/nips/InalaWYCELMG22, DBLP:conf/icml/ZhangYHLYF023, DBLP:conf/emnlp/ShiFGZW22}, natural language self-critique~\citep{DBLP:conf/acl/ZhangLLLJ23}, and reasoning about compiler feedback~\citep{DBLP:conf/iclr/ChenLSZ24}.
Alternatively, prior work has sought to train LLMs with execution semantics to directly~\citep{zhu2026codescalerscalingcodellm,DBLP:conf/icml/NiACDSSY24} or indirectly~\citep{DBLP:journals/corr/abs-2510-02387, DBLP:journals/corr/abs-2509-22824} improve their ability to abstractly reason about code execution. Beyond the file level, prior work has sought to reason about repository-level test-suite execution outcomes for software engineering tasks~\citep{Shum2025SWERMEF, DBLP:conf/icml/Pan0NJ0S025}.
In this work, we show that RLVR enables training robust code verifiers that can scalably supervise much larger policy models, and that using a subset of its core components can even be optimal, yielding competitive results.

\paragraph{Prior analyses of RLVR in LLM training.}
Modern RL training is notoriously compute-intensive and inefficient~\citep{DBLP:conf/iclr/NoukhovitchHXHA25,DBLP:journals/corr/abs-2509-19128}, prompting a surge of work simplifying the RL recipe by omitting core components like long thinking traces~\citep{arora2025traininglanguagemodelsreason,sui2025stopoverthinkingsurveyefficient}, negative samples~\citep{dongRAFTRewardRAnked, gulcehreReinforcedSelfTrainingReST2023, singhHumanDataScaling2024}, and on-policy learning
~\citep{rafailovDirectPreferenceOptimization2024, wangCritiqueFineTuningLearning2025}. However, the contribution of these components to RLVR's success is unclear, as described in \cref{sec: think-toggle,sec: online-toggle,sec: neg-toggle}. Moreover, they are often treated as isolated rather than synergistic contributors to RLVR's success. A notable exception is ~\citet{tajwar2024preference}, who found that on-policy learning and negative samples are complementary and especially important when high-reward responses are less likely under the policy distribution. Additionally, most RLVR studies focus on mathematical reasoning. It is unclear whether these findings transfer to more brittle domains like code verification, where even frontier models frequently fail~\citep{DBLP:journals/corr/abs-2504-04372, DBLP:journals/corr/abs-2502-11167}. In this work, we deconstruct the training dynamics of code verifiers to reveal that the importance of specific components varies with scale. We provide a roadmap for efficiently training code verifiers to supervise future generations of code models.

\section{Conclusion}
\label{sec: conclusion}
In this work, we presented a systematic analysis of the three primary drivers of performance and cost in the RLVR pipeline for code verifiers: generating intermediate thinking traces, learning from positive and negative samples, and on-policy training. To facilitate this study, we introduced \textbf{\workname{}}, a controlled, execution-grounded testbed designed to draw decontaminated conclusions about the training dynamics of code verifiers across different model sizes and covariate shifts with proxy metrics for two common verifier application scenarios: Best-of-N inference and RL reward modeling.
Our analysis reveals that these components are synergistic, but the degree of their impact is scale-dependent: for both top-1 selection and reranking, on-policy learning is the primary performance driver for small verifiers, while thinking becomes the most vital factor as model size increases. Negatives consistently boost performance on top-1 selection, are monotonic contributors to reranking performance, and prevent reward curves from degrading at larger sizes. We find that scaling inference-time compute with self-consistency yields only a minor performance boost in most cases and cannot compensate for the absence of any component. Finally, we conduct a Pareto optimality analysis of our verifiers and find that \texttt{DPO-Think-14B} is an attractive choice for training verifiers when a substantial decrease in cost is worth more than a small drop in performance. Although the full RLVR recipe is more performant, its cost is justified only when shifts in generator capability and adversarial perturbations are expected. At low budgets, \texttt{GRPO-Instruct-7B} is a strong baseline and is optimal across all evaluations, similar to \texttt{DPO-Think-14B}. We observe our findings are well-supported under a downstream best-of-N integration experiment as well as on two external benchmarks. Therefore, our work establishes a practical cost\,--\,performance guide for practitioners by providing strategies to simplify verifier training across several scales and analyzing the consequences of these simplifications across multiple covariate shifts. More broadly, our work lays the foundation for generative code verifiers to become a more prominent fixture in the post-training pipelines of code generator LLMs by providing recipes for efficient verifier training.

\section*{Limitations}
\label{sec: limitations}
Our work studies surrogate verifiers for code generation, an area that remains underexplored compared to standard reward modeling. Code contests offer a unique advantage for our research due to large open-source datasets providing reliable and verifiable ground truth signals with limited setup overheads. However, the true utility of such a verifier is in domains that are harder to verify via execution. Creating a controlled and execution-grounded testbed like ours for such domains is a significant challenge due to environment setup and verification overheads, and could be a potential direction for future work. Another limitation of our work is that truly isolating each RLVR component is difficult, and might introduce some confounding factors into our analysis. We mitigate the effects of such factors by carefully curating our offline DPO dataset with the same problems and generators as encountered in the online setting and equally partitioning negative samples across the three verifier sizes to avoid biases (\cref{sec: detailed-hparams}). We also ablate multiple RAFT variants in \cref{sec: extended-raft} and report the best performing one in our main results. Our results may also be sensitive to the model family used. Although we validate our findings on \texttt{Olmo3-7B} in \cref{sec: olmo-results}, we were unable to extend this analysis to more model families and sizes due to limited computational resources. We were also unable to validate our findings across multiple random seed settings due to the same constraints. Lastly, while we take steps to ensure our testbed is free from data contamination, one cannot guarantee that the coding problems have not been encountered in earlier training stages. However, the responses to the coding problems are self-generated and mitigates this issue to an extent. Moreover, we do see a considerable drop in performance in our OOD settings, indicating that the effects of any potential contamination is minimal.

\section*{Ethical Considerations}
\label{sec: risks}
Our work focuses on training recipes for code verifiers that are used to select the best code from a list of LLM-generated snippets. These verifiers can potentially be exploited by an adversary to generate incorrect or unsafe code. We take steps to mitigate these risks by analyzing the OOD robustness of our verifiers under various shifts in generator capability and in adversarial settings, and report our findings on these evaluations separately. To further encourage research in this area, we thoroughly document our workflow and open-source our datasets, code, and models under the \texttt{CC BY-NC-SA 4.0} License \faCreativeCommons\ \faCreativeCommonsBy\ \faCreativeCommonsNc\ \faCreativeCommonsSa.

\section*{Acknowledgements}
This research was partially funded by the Ministry of Education and Science of Bulgaria (support for INSAIT, part of the Bulgarian National Roadmap for Research Infrastructure). Additionally, we gratefully acknowledge the support from (1) the hessian.AI Service Center (funded by the Federal Ministry of Research, Technology and Space, BMFTR, grant no. 16IS22091), (2) the hessian.AI Innovation Lab (funded by the Hessian Ministry for Digital Strategy and Innovation, grant no. S-DIW04/0013/003), (3) the German Federal Ministry of Research, Technology, and Space and the Hessian Ministry of Higher Education, Research, Science, and the Arts within their joint support of the National Research Center for Applied Cybersecurity ATHENE, and (4) computational resources provided by the Google Cloud Platform (GCP).
\bibliography{custom}
\bibliographystyle{tmlr}

\clearpage
\tableofcontents
\appendix

\section{Additional Experiment Details}
\label{sec: detailed-hparams}
We use the implementation provided by \citet{xiongMinimalistApproachLLM2025} for \texttt{RAFT} and the \texttt{trl} library\footnote{\href{https://github.com/huggingface/trl}{\scalerel*{\includegraphics{Figures/github_logo.png}}{\textbf{C}} \texttt{huggingface/trl}}} for \texttt{GRPO} and \texttt{DPO-Think}. All training runs are conducted on a cluster of \texttt{8 NVIDIA H200} GPUs. To optimize memory usage, we employ Deepspeed ZeRO Stage-2~\citep{rasley2020deepspeed} to shard activations and optimizer states across devices, and Flash-Attention 3~\citep{shah2024flashattention3fastaccurateattention} to accelerate training. In all our training runs, we use the AdamW optimizer~\citep{loshchilov2019decoupledweightdecayregularization} with default parameters and a constant learning rate scheduler with \texttt{5\%} warmup steps, and train with an effective batch size of \texttt{64} for exactly \texttt{781} gradient steps to ensure a fair comparison.

Our \texttt{GRPO} implementation deviates from the original~\citep{deepseekai2025deepseekr1incentivizingreasoningcapability} to incorporate future recipe refinements. We use the DAPO loss~\citep{yuDAPOOpenSourceLLM2025} and Truncated Importance Sampling~\citep{yao2025offpolicy} with the truncation threshold set to \texttt{2.0}. Although recent works have chosen to eliminate the KL coefficient, we set it to $\beta$\,=\,\texttt{1e-3} because our base models are already fine-tuned to generate long reasoning traces. We synchronize the reference model every \texttt{100} steps~\citep{gorbatovski2024learn, liuProRLProlongedReinforcement2025}. We use a learning rate of \texttt{1e-6} and normalize by the standard deviation within each group. We note that while \citet{liuPartTricksTraps2025} suggest batch-level normalization for base models, their results indicate poor performance for aligned models, such as those used in this study. We generate a batch of \texttt{64} prompts and perform a single gradient update per batch, with $\epsilon_\text{low}$\,=\,\texttt{0.2} and $\epsilon_\text{high}$\,=\,\texttt{0.28}. To encourage the model to stay within budget, we use a soft overlong punishment reward~\citep{yuDAPOOpenSourceLLM2025}.

For \texttt{BO-GRPO}, we use a generation batch of \texttt{256} prompts, performing \texttt{4} gradient updates per batch with $\epsilon_\text{low}$\,=\,\texttt{3e-4}, $\epsilon_\text{high}$\,=\,\texttt{4e-4} and sequence-level importance sampling~\citep{zhengGroupSequencePolicy2025}. All other details are the same as the online \texttt{GRPO} variant.

Following~\citet{lambertTulu3Pushing2025}, we train \texttt{DPO} with a learning rate of \texttt{5e-7}, KL penalty $\beta$\,=\,\texttt{0.1}, and an effective training batch size of \texttt{64}. To reduce memory overhead, we precompute log-probabilities, eliminating the need to load the reference model during training. To train a \texttt{DPO} model, we also need an offline dataset of preferred and dispreferred generations. To this end, we create \dpo{} by sampling \texttt{100} outputs for each prompt in \train{} using \texttt{DeepSeek-R1-Distill-Qwen-[1.5-14]B}, and score them with our verifiable reward function.

While prior work finds the quality of chosen responses to be more important~\citep{panWhatMattersData2025}, we hypothesize that the reverse is true in a verifiable setting, where the quality of the ``chosen'' sample is fixed (correct), but the rejected quality can vary. Moreover, DPO is known to be sensitive to OOD shifts~\citep{xuDPOSuperiorPPO2024}. Thus, we distribute the incorrect responses evenly between those generated by the \texttt{1.5-14B} models. This also ensures that the negative samples for \texttt{DPO} are drawn from generations similar to those from on-policy sampling. Our hypothesis is validated by the strong performance of our \texttt{DPO} models, even rivaling the fully online \texttt{GRPO} at larger sizes.

\texttt{RAFT} is trained with a learning rate of \texttt{2e-6} and an effective batch size of \texttt{64}. Consistent with~\citet{dongRAFTRewardRAnked, xiongMinimalistApproachLLM2025}, no KL penalty is applied. In preliminary runs, we found that fine-tuning on the entire batch of correct responses leads to overfitting, especially in large models that generate a high proportion of correct responses. We mitigate this effect by fine-tuning on a maximum of \texttt{5} correct responses per group (See \cref{sec: extended-raft} for more details).
\section{Alternate Reward Formulations}
\label{sec: alt-rewards}
Shaping the reward during RL training is a crucial decision, and numerous proposals for optimal reward functions have been made in prior work. We experiment with four reward formulations at \texttt{7B} model scale and pick the best-performing one for our final training runs. The rewards used are as follows:
\begin{itemize}[leftmargin=*, noitemsep]
    \item \textbf{Pairwise Exact Match (PairEM)}. The simplest formulation. Given two candidates, we prompt the verifier to indicate its preference with a single token (A or B) within $\texttt{\\boxed\{\}}.$
    \item \textbf{Pairwise Scores (PairSc)}. This reward is taken from the JudgeLRM paper~\citep{DBLP:journals/corr/abs-2503-01307}. The verifier outputs scores on a scale of \texttt{0-10} for both candidate codes, and the reward is shaped based on accuracy, confidence, and format.
    \item \textbf{Listwise Exact Match (ListEM)}. A modified version of PairEM with between two and five candidates
    \item \textbf{Listwise Scores (ListSc)}. The verifier outputs scores on a \texttt{10}-point scale for each candidate, similar to PairSc. If the correct code is assigned the highest score, we assign a reward of \texttt{+1} and a bonus of \texttt{+1} if this score is \texttt{10}.
\end{itemize}
Both listwise rewards are loosely based on DeepSeek-GRM~\citep{liu2025inferencetimescalinggeneralistreward}, adapted to our setting. For PairSc and ListSc, we use the pass rate of both codes as an indication of their quality. Since one of the codes is always correct, one of the scores outputted by the model should always be \texttt{10}. We train these models using GRPO as described in the main paper and present the results in \cref{tab: alternate-rewards}.
\begin{table}[ht]
    \centering
    \caption{\textbf{Average List Accuracy results for alternate reward formulations we studied.} All results are from training the \texttt{7B} model for an equal number of gradient updates using \texttt{GRPO}. For a fair comparison, we evaluate on code pairs, which explains the higher absolute values compared to \cref{sec: results}.}
    \small
    \tt
    \scalebox{0.8}{
        \begin{tabular}{l|cccc}
            \toprule
            \textbf{Reward} & \textbf{ListAcc@1}             & \textbf{ListAcc@2}             & \textbf{ListAcc@4}             & \textbf{ListAcc@8}             \\
            \midrule
            PairSC          & {78.24\,$_\text{0.93}$}        & {78.13\,$_\text{0.93}$}        & {80.16\,$_\text{1.00}$}        & {82.22\,$_\text{1.06}$}        \\
            PairEM          & {77.19\,$_\text{0.90}$}        & {77.50\,$_\text{0.90}$}        & {78.93\,$_\text{0.95}$}        & {80.12\,$_\text{0.10}$}        \\
            ListSC          & {77.36\,$_\text{0.89}$}        & {77.32\,$_\text{0.89}$}        & {79.40\,$_\text{0.94}$}        & {80.82\,$_\text{0.98}$}        \\
            ListEM          & \textbf{80.02\,$_\text{0.92}$} & \textbf{80.14\,$_\text{0.92}$} & \textbf{81.50\,$_\text{0.94}$} & \textbf{83.02\,$_\text{0.98}$} \\
            \bottomrule
        \end{tabular}
    }
    \label{tab: alternate-rewards}
\end{table}
We find that relatively simple ListEM works best, followed by PairSc. Moreover, we find that inference-time scaling trends hold even for these alternative reward formulations, providing limited benefits. This is also consistent with prior work that finds RLVR tends to sharpen the output distribution of models~\citep{yueDoesReinforcementLearning2025}.
\section{Modifications for \rqfour{}}
\label{sec: adversarial-mods}
\begin{table}[!ht]
    \centering
    \caption{
        \textbf{Modifications considered to construct \rqfour{}.}
        We report the Bias Influence Ratio (BIR) for the \texttt{7\,--\,32B} models, along with the average. Positive modifications are applied to the incorrect code, whereas negative ones are applied to the correct one. The top six modifications are highlighted.
    }
    \tt
    \small
    \scalebox{0.8}{
        \begin{tabular}{l|>{\raggedright\arraybackslash}p{11cm}|cccc}
            \toprule
            \textbf{Name}                        & \textbf{Description}                                                                              & \textbf{7B}   & \textbf{14B}  & \textbf{32B}  & \textbf{Avg.} \\
            \midrule
            \multicolumn{6}{c}{Positive Biases}                                                                                                                                                                      \\
            \midrule
            \textbf{Authority Bias}              & Claims the incorrect code was written by an experienced developer.                                & \textbf{0.56} & \textbf{0.67} & \textbf{0.67} & \textbf{0.64} \\
            Egocentric bias                      & Indicates an incorrect code was written by the evaluator                                          & 0.52          & 0.49          & 0.54          & 0.52          \\
            \textbf{External Reference}          & Claims to be the reference solution on the competition's website                                  & \textbf{0.58} & \textbf{0.78} & \textbf{0.85} & \textbf{0.73} \\
            Bandwagon Effect                     & Indicates that a majority of developers prefer the incorrect code.                                & 0.51          & 0.55          & 0.55          & 0.54          \\
            Illusory Complexity                  & Garbage/unreachable code to elicit length bias~\citep{zheng2023judgingllmasajudgemtbenchchatbot}. & 0.40          & 0.44          & 0.49          & 0.44          \\
            \textbf{Self-declared correctness}   & States that the code is correct                                                                   & \textbf{0.64} & \textbf{0.77} & \textbf{0.75} & \textbf{0.72} \\
            \midrule
            \multicolumn{6}{c}{Negative Biases}                                                                                                                                                                      \\
            \midrule
            Minification                         & Code compressed using a rule-based minifier for C++ and Java,
            and \texttt{python-minifier}\footnote{\href{https://github.com/dflook/python-minifier}{\scalerel*{\includegraphics{Figures/github_logo.png}}{\textbf{C}} \texttt{dflook/python-minifier}}}
            for Python.                          & 0.50                                                                                              & 0.52          & 0.50          & 0.51                          \\
            \textbf{Misleading Comments}         & Comments indicate the correct code makes an error                                                 & \textbf{0.53} & \textbf{0.76} & \textbf{0.82} & \textbf{0.71} \\
            Renaming Identifiers                 & Variable, class, and function names are obfuscated~\citep{paul2025obscuracoder}                   & 0.54          & 0.60          & 0.54          & 0.56          \\
            \textbf{Reverse Authority Bias}      & Claims the incorrect code was written by a junior developer                                       & \textbf{0.53} & \textbf{0.71} & \textbf{0.65} & \textbf{0.63} \\
            Reverse Bandwagon Effect             & Indicates that a minority of developers prefer the correct code.                                  & 0.44          & 0.60          & 0.56          & 0.53          \\
            \textbf{Self-declared incorrectness} & States that the code is incorrect                                                                 & \textbf{0.60} & \textbf{0.81} & \textbf{0.86} & \textbf{0.76} \\
            \bottomrule
        \end{tabular}
    }
    \label{tab: mods-switching}
\end{table}

We experiment with several biasing factors for the creation of \rqfour{} (\cref{tab: mods-switching}). To analyze the vulnerability of the base models to these factors, we prompt \texttt{Deepseek-R1-Distill-Qwen2.5} from \texttt{7B} to \texttt{32B} parameters on the original and perturbed versions of the same prompt, and measure how often the evaluator switches its answer. Positive modifications are applied to all incorrect codes, while negative modifications are applied only to the correct code.

We conduct evaluations of perturbation effectiveness in a pairwise setting. We report the Bias Influence Ratio (BIR) as the ratio of the number of times the LLM switches to the incorrect answer to the total number of switches. A higher BIR indicates a stronger bias, while a BIR near \texttt{0.5} indicates random answer switching, which could be due to position biases in the model~\citep{zheng2023judgingllmasajudgemtbenchchatbot}. Overall, we verify that LRMs are more robust to common biases that are prevalent in LLMs, as observed in prior work~\citep{wangAssessingJudgingBias2025}. However, they are not completely unbiased, and we select the top six most misleading modifications for analysis of adversarial robustness in the main text.

\section{Alternative Approaches to Implement RAFT}
\label{sec: extended-raft}
While designing our Negatives ablation in \cref{sec: neg-toggle}, we experimented with several variants of the RAFT algorithm. A crucial requirement was that our RAFT algorithm be fully on-policy to isolate the effect of negative samples. The original RAFT~\citep{dongRAFTRewardRAnked}, akin to common rejection-sampling algorithms, is batch-online: sampling $N$ responses each for a batch of prompts from the current model, scoring them using a reward model, and fine-tuning the current model on the $K$ highest scoring prompt-response pairs using a negative log-likelihood loss. Moreover, this algorithm was designed for RLHF-style reward models that output a continuous scalar score, unlike the binary reward signals prevalent in GRPO.

Our implementation closely follows \citet{xiong2023iterative}, who adapt RAFT to the RLVR setting, with some modifications. Most importantly, we collect new data after a single gradient update (which slows training but ensures it remains fully on-policy). An unclear implementation detail in their work is whether they train on \textit{all} the correct responses for a group or only a subset. We experiment with both variants, and the RAFT$++$ algorithm proposed by \citet{xiong2023iterative}, on the \texttt{14B} model as shown in \cref{tab: raft-variants}.
\begin{table}[h]
    \centering
    \caption{\textbf{Experimental results for RAFT variants.} We present the average \texttt{ListAcc@1} scores across the four \texttt{Aletheia-} evaluation datasets, along with their $95\%$ confidence interval.}
    \small
    \tt
    \scalebox{0.85}{
        \begin{tabular}{l|c|p{0.6\linewidth}}
            \toprule
            \textbf{Algorithm}          & \textbf{ListAcc@1}        & \textbf{Description}                                                         \\ \midrule
            \texttt{RAFT}               & 51.20 $\pm$ 0.53          & \texttt{RAFT} adapted to verifiable rewards, trained on all positives        \\
            \textbf{\texttt{RAFT-max5}} & \textbf{67.20 $\pm$ 0.52} & \texttt{RAFT} trained on a maximum of \texttt{5} correct responses per group \\
            \texttt{RAFT++}             & 60.41 $\pm$ 0.47          & Variant adding importance sampling and clipping to \texttt{RAFT}             \\ \bottomrule
        \end{tabular}
    }
    \label{tab: raft-variants}
\end{table}

Clearly, fine-tuning on all correct responses for each group leads to overfitting on easy examples, yielding a higher number of correct responses for training than for harder prompts. Surprisingly, adding PPO-style importance sampling and clipping techniques does not stabilize training either, further emphasizing the role of negative samples. We use the best performing max-\texttt{5} variant of \texttt{RAFT} for all results presented in the main text.
\section{Supporting Results}
\label{sec: supporting-results}
\subsection{Supervised Fine-tuning}
\label{sec: sft-results}
We train models using the Supervised Fine-Tuning (SFT) objective on the positively scored responses from \dpo{} (as described in \cref{sec: detailed-hparams}) across the \texttt{1.5-14B} scale to further validate our analyses of the Negatives and Online components. Notably, we omit this algorithm from the main text because it differs from \texttt{GRPO} along two axes: Negatives and Online. However, in this section, we compare it to the other ``incomplete'' algorithms: \texttt{DPO-Think}, \texttt{BO-GRPO}, and \texttt{RAFT}. We report results in \cref{tab: sft-results}.

\begin{table*}[!t]
        \centering
        \caption{\textbf{Additional results on an SFT baseline.} We report \texttt{ListAcc} scores at \texttt{K=1}. SFT performs the worst overall due to the absence of two RLVR components. At larger scales, on-policy learning is the least critical factor, and negatives play a critical role in training stability for all sizes. Subscripts are 95\% confidence intervals.}
        \small
        \tt
        \setlength{\tabcolsep}{6pt}\renewcommand{\arraystretch}{1.3}
        \scalebox{0.8}{
                \begin{tabular}{clc|c|c|c|c|c}
                        \multicolumn{8}{c}{\thinking{}~\texttt{\textbf{Thinking}} \quad \negatives{}~\texttt{\textbf{Negatives}} \quad \online{}~\texttt{\textbf{Online}} \quad \semionline{}~\texttt{\textbf{Batch-online}}} \\
                        \toprule
                                                               & \textbf{Method} & \textbf{Size}         & \textbf{\heldout{}}  & \textbf{\rqone{}}    & \textbf{\rqtwo{}}    & \textbf{\rqfour{}}   & \textbf{Average}       \\ \midrule
                        -                                      & Random          &                       & \scval{32.08}        & \scval{32.08}        & \scval{32.08}        & \scval{32.08}        & \scval{32.08}          \\\midrule
                        \thinking{} \nonegatives{} \noonline{} & SFT-Think       & \multirow{5}{*}{1.5B} & \scvalc{20.27}{0.43} & \scvalc{18.23}{0.45} & \scvalc{18.30}{0.47} & \scvalc{18.00}{0.14} & \scvalc{18.70}{0.37}   \\
                        \thinking{} \negatives{} \noonline{}   & DPO-Think       &                       & \scvalc{21.70}{0.39} & \scvalc{19.90}{0.35} & \scvalc{23.41}{0.37} & \scvalc{19.37}{0.15} & \scvalc{21.11}{0.32}   \\
                        \thinking{} \negatives{} \semionline{} & BO-GRPO         &                       & \scvalc{43.13}{0.55} & \scvalc{39.89}{0.52} & \scvalc{39.26}{0.54} & \scvalc{33.83}{0.22} & \scvalc{38.99}{0.46}   \\
                        \thinking{} \nonegatives{} \online{}   & RAFT            &                       & \scvalc{34.76}{0.48} & \scvalc{31.88}{0.44} & \scvalc{33.67}{0.46} & \scvalc{29.12}{0.19} & \scvalc{32.30}{0.39}   \\
                        \thinking{} \negatives{} \online{}     & GRPO-Think      &                       & \scvalc{49.58}{0.65} & \scvalc{46.09}{0.63} & \scvalc{40.74}{0.64} & \scvalc{40.97}{0.26} & \scvalc{44.38}{0.55}   \\\midrule
                        \thinking{} \nonegatives{} \noonline{} & SFT-Think       & \multirow{5}{*}{7B}   & \scvalc{46.70}{0.52} & \scvalc{40.30}{0.54} & \scvalc{31.40}{0.61} & \scvalc{41.11}{0.24} & \scvalc{39.87}{0.48}   \\
                        \thinking{} \negatives{} \noonline{}   & DPO-Think       &                       & \scvalc{63.75}{0.57} & \scvalc{55.54}{0.57} & \scvalc{51.20}{0.62} & \scvalc{52.94}{0.25} & \scvalc{55.88}{0.50}   \\
                        \thinking{} \negatives{} \semionline{} & BO-GRPO         &                       & \scvalc{64.71}{0.60} & \scvalc{56.18}{0.61} & \scvalc{52.08}{0.66} & \scvalc{49.35}{0.27} & \scvalc{55.46}{0.54}   \\
                        \thinking{} \nonegatives{} \online{}   & RAFT            &                       & \scvalc{60.86}{0.59} & \scvalc{52.00}{0.58} & \scvalc{48.84}{0.63} & \scvalc{49.24}{0.26} & \scvalc{52.72}{0.52}   \\
                        \thinking{} \negatives{} \online{}     & GRPO-Think      &                       & \scvalc{74.81}{0.57} & \scvalc{67.28}{0.60} & \scvalc{53.11}{0.69} & \scvalc{65.04}{0.26} & \scvalc{65.05}{0.53}   \\ \midrule
                        \thinking{} \nonegatives{} \noonline{} & SFT-Think       & \multirow{5}{*}{14B}  & \scvalc{66.53}{0.50} & \scvalc{60.30}{0.53} & \scvalc{48.53}{0.67} & \scvalc{59.04}{0.26} & \scvalc{58.60}{0.49}   \\
                        \thinking{} \negatives{} \noonline{}   & DPO-Think       &                       & \scvalc{82.56}{0.52} & \scvalc{74.39}{0.58} & \scvalc{67.58}{0.68} & \scvalc{71.06}{0.26} & \scvalc{73.89}{0.51}   \\
                        \thinking{} \negatives{} \semionline{} & BO-GRPO         &                       & \scvalc{83.82}{0.50} & \scvalc{76.33}{0.56} & \scvalc{67.34}{0.68} & \scvalc{73.45}{0.25} & \scvalc{75.29}{0.50}   \\
                        \thinking{} \nonegatives{} \online{}   & RAFT            &                       & \scvalc{75.55}{0.56} & \scvalc{66.02}{0.61} & \scvalc{65.23}{0.65} & \scvalc{62.03}{0.27} & \scvalc{67.20}{0.52}   \\
                        \thinking{} \negatives{} \online{}     & GRPO-Think      &                       & \scvalc{88.02}{0.45} & \scvalc{83.65}{0.49} & \scvalc{66.84}{0.70} & \scvalc{83.67}{0.21} & \scvalc{80.54}{0.46}   \\ \bottomrule
                \end{tabular}
        }
        \label{tab: sft-results}
\end{table*}

Removing two of the core components from \texttt{GRPO} significantly hampers performance across all model sizes. Our observations about the importance of on-policy learning decreasing with scale remain valid, as evidenced by the \texttt{SFT}--\texttt{RAFT} gap decreasing with scale (from \texttt{13.6\%} to \texttt{8.6\%}). Crucially, the utility of negative samples for stabilizing training is more pronounced in these ablations, as shown by comparing the \texttt{SFT}--\texttt{RAFT} gap to the \texttt{SFT}--(\texttt{BO-GRPO}) gap. Despite \texttt{BO-GRPO} being only partially on-policy, it outperforms the fully on-policy \texttt{RAFT} across all model scales due to the presence of negative samples. At small scales, using \texttt{BO-GRPO}: a mixture of on-policy training and negatives, is the best alternative to the full \texttt{GRPO} algorithm. However, at medium--large scales, using a completely offline algorithm (\texttt{DPO-Think}) can already yield good results at much lower cost, as mentioned in \cref{sec: analysis}.

\subsection{Training on a Mixed Dataset}
\label{sec: mixed-data}
A key advantage of the \workname{} testbed is the complete separation of training and evaluation data distributions, enabling an accurate estimation of the OOD robustness of our trained verifiers. However, this setting differs from the practical approach: training verifiers on a mix of all anticipated scenarios to achieve the best downstream performance. For completeness, we present the results from training a \texttt{1.5B} model on a mixture of all our evaluation scenarios. Concretely, we utilized unused instances from the \weak{}-\hard{} and \sstrong{}-\easy{} buckets (i.e., instances that do not appear in the corresponding evaluation sets), and created a new adversarial dataset by perturbing \train{} using the six best modifications detailed in \cref{sec: adversarial-mods}. Crucially, we now apply the modifications at random rather than targeting either the correct or incorrect code, thereby avoiding information leakage about the ground truth and training the verifier to be robust to such perturbations. The resulting \mixed{} contains all four data distributions in equal proportions. We create a corresponding DPO dataset following the same procedure as in \cref{sec: detailed-hparams}, and summarize our results in \cref{tab: mixed-evals}.

\begin{table}[!t]
        \centering
        \caption{\textbf{Results from training \texttt{1.5B} verifiers on \mixed{}.} We report \texttt{ListAcc} scores at \texttt{K=1}. Training on a mixed dataset yields minor performance boosts over the \workname{} testbed in most cases, but sacrifices conclusions about OOD robustness. Subscripts are 95\% confidence intervals.}
        \small
        \tt
        \setlength{\tabcolsep}{6pt}\renewcommand{\arraystretch}{1.3}
        \scalebox{0.8}{
                \begin{tabular}{ll|ccc}
                        \multicolumn{5}{c}{\thinking{}~\texttt{\textbf{Thinking}} \quad \negatives{}~\texttt{\textbf{Negatives}} \quad \online{}~\texttt{\textbf{Online}} \quad \semionline{}~\texttt{\textbf{Batch-online}}} \\
                        \toprule
                                                               & \textbf{Algorithm} & \textbf{\rqone{}}    & \textbf{\rqtwo{}}    & \textbf{\rqfour{}}                                                                        \\ \midrule
                        \nothinking{} \negatives{} \online{}   & GRPO-Instruct      & \scvalc{37.11}{0.87} & \scvalc{29.68}{0.82} & \scvalc{31.61}{0.34}                                                                      \\
                        \thinking{} \negatives{} \noonline{}   & DPO-Think          & \scvalc{20.37}{0.39} & \scvalc{24.24}{0.43} & \scvalc{20.08}{0.17}                                                                      \\
                        \thinking{} \negatives{} \semionline{} & BO-GRPO            & \scvalc{40.11}{0.54} & \scvalc{39.19}{0.54} & \scvalc{33.42}{0.23}                                                                      \\
                        \thinking{} \nonegatives{} \online{}   & RAFT               & \scvalc{32.99}{0.49} & \scvalc{34.06}{0.50} & \scvalc{29.67}{0.21}                                                                      \\
                        \thinking{} \negatives{} \online{}     & GRPO-Think         & \scvalc{41.46}{0.58} & \scvalc{39.05}{0.57} & \scvalc{34.09}{0.24}                                                                      \\ \bottomrule
                \end{tabular}
        }
        \label{tab: mixed-evals}
\end{table}
Training on a mixture of all tasks yields small performance gains for most algorithms, as compared to the results in \cref{tab: think-toggle-merged,tab: online-toggle-merged,tab: neg-toggle-merged}. More importantly, our conclusions from the main text at the \texttt{1.5B} scale still hold. \texttt{DPO} performs the worst at this scale, and introducing even a semi-on-policy update can close most of the offline-online performance gap, making on-policy learning the most crucial component at this scale.

Crucially, training on a mixed dataset precludes analysis of the OOD robustness of the studied algorithms. Since \mixed{} contains a mixture of all anticipated evaluation scenarios, all three evaluation axes are in-distribution, providing no signal on the robustness of these verifiers in a downstream RLVR pipeline. Thus, by completely separating the training and evaluation data distributions, the \workname{} testbed provides a foundation for our controlled analysis, enabling us to stress-test verifiers in a proxy-evaluation setting without incurring prohibitive costs.

\subsection{Inference-time scaling on \texttt{RunBugRun}}
\label{sec: inference-comp}
To ensure the validity of our inference-time scaling observations across reward formulations, we further test our hypothesis on the \texttt{RunBugRun} dataset~\citep{runbugrun}, which contains correct-bugged code pairs from CodeNet~\citep{DBLP:conf/nips/Puri0JZDZD0CDTB21}. We report our results in \cref{tab: inf-rbr}.

\begin{table}[h]
        \centering
        \caption{\textbf{Inference scaling trends on the \texttt{RunBugRun} dataset.} We report the \texttt{ListAcc} scores for different reward formulations`'. \texttt{GRPO}-trained models see limited gains from inference compute scaling. Subscripts are 95\% confidence intervals.}
        \small
        \tt
        \setlength{\tabcolsep}{6pt}\renewcommand{\arraystretch}{1.3}
        \scalebox{0.8}{
                \begin{tabular}{lcccc}
                        \toprule
                        \textbf{Reward} & \textbf{RBR PairAcc@1}  & \textbf{RBR PairAcc@2}  & \textbf{RBR PairAcc@4}  & \textbf{RBR PairAcc@8}  \\ \midrule
                        PairSC          & 67.43\ci{1.12}          & 67.23\ci{1.12}          & 69.48\ci{1.21}          & 71.08\ci{1.30}          \\
                        PairEM          & 70.84\ci{1.14}          & 70.83\ci{1.13}          & 72.99\ci{1.22}          & 73.61\ci{1.30}          \\
                        ListSC          & 71.01\ci{1.21}          & 71.26\ci{1.21}          & 73.12\ci{1.29}          & 74.17\ci{1.37}          \\
                        ListEM          & \textbf{73.24}\ci{1.14} & \textbf{73.63}\ci{1.13} & \textbf{74.99}\ci{1.22} & \textbf{75.61}\ci{1.30} \\ \bottomrule
                \end{tabular}
        }
        \label{tab: inf-rbr}
\end{table}

The modest gains from inference-time scaling persist across an external dataset, demonstrating that our observation is a consistent pattern rather than an artifact of our specific setup. Moreover, it is supported by prior literature on entropy collapse in LLMs, which finds that RLVR merely improves the sampling efficiency of language models and can shrink the space of accessible reasoning paths~\citep{yueDoesReinforcementLearning2025,wu2025invisible}.

\subsection{Verifier Response Parseability}
\label{sec: parse-rates}
The parse rate of a verifier is defined as the percentage of its responses that terminate with a valid and parseable verdict. In all the verifiers studied in the main text, this implies that the verifier responds with a valid option within \texttt{$\backslash$boxed\{\}}. The models are trained to respond with such a format, either through demonstrations or via a format reward function. We report the parse rates for all models in \cref{tab: parse-rates}. As expected, most models achieve near-perfect parse rates, since they are being explicitly trained to do so.

However, a notable exception is \texttt{DPO-Think-1.5B}, which has a very poor parse rate of only \rcval{43\%} on average. This is because the model degenerates into producing non-terminating sequences. Therefore, such a verifier has very limited utility in practice despite seemingly high \ktau{} scores. These high scores are simply explained by the fact that \ktau{} effectively treats unparseable responses as a tie between the two candidates, rather than as a failure mode of the verifier as in \texttt{ListAcc} (which expectedly degrades to sub-random performance). This degeneration is likely an artifact of the small model size rather than the offline dataset, since the larger \texttt{DPO-Think} verifiers have high parse rates. Thus, we suggest against using \ktau{} alone as a measure of verifier performance, and instead augment it with a parse rate check.
\begin{table}[h]
    \centering
    \caption{\textbf{Parse rates for the algorithms studied in this work.} \texttt{DPO-Think-1.5B} yields an abnormaly low number of parseable verdicts, but this issue is not present in the larger models.}
    \small
    \tt
    \scalebox{0.8}{
        \begin{tabular}{llcccccc}
            \multicolumn{8}{c}{\thinking{}~\textbf{Thinking} \quad \negatives{}~\textbf{Negatives} \quad \online{}~\textbf{Online} \quad \semionline{}~\textbf{Batch-online}}                                        \\
            \toprule
                                                   & \textbf{Algorithm}                        & \textbf{Size} & \textbf{\heldout{}} & \textbf{\rqone{}} & \textbf{\rqtwo{}} & \textbf{\rqfour{}} & \textbf{Average} \\ \midrule
            \nothinking{} \negatives{} \online{}   & (\S\ref{sec: think-toggle}) GRPO-Instruct &               & \miscval{99.97}     & \miscval{100.00}  & \miscval{100.00}  & \miscval{100.00}   & \miscval{99.99}  \\
            \thinking{} \negatives{} \online{}     & (\S\ref{sec: think-toggle}) GRPO-Think-4k &               & \miscval{99.78}     & \miscval{99.72}   & \miscval{99.14}   & \miscval{99.30}    & \miscval{99.49}  \\
            \thinking{} \negatives{} \online{}     & (\S\ref{sec: think-toggle}) GRPO-Think-8k &               & \miscval{99.14}     & \miscval{99.54}   & \miscval{98.55}   & \miscval{99.40}    & \miscval{99.16}  \\
            \thinking{} \negatives{} \noonline{}   & (\S\ref{sec: online-toggle}) DPO-Think    & 1.5B          & \miscval{40.10}     & \miscval{41.15}   & \miscval{49.01}   & \miscval{42.64}    & \miscval{43.23}  \\
            \thinking{} \negatives{} \semionline{} & (\S\ref{sec: online-toggle}) BO-GRPO      &               & \miscval{98.98}     & \miscval{98.80}   & \miscval{97.81}   & \miscval{98.73}    & \miscval{98.58}  \\
            \thinking{} \nonegatives{} \online{}   & (\S\ref{sec: neg-toggle}) RAFT            &               & \miscval{71.50}     & \miscval{74.58}   & \miscval{75.48}   & \miscval{72.03}    & \miscval{73.40}  \\
            \thinking{} \negatives{} \online{}     & (\S\ref{sec: results})~~~GRPO-Think-16k   &               & \miscval{99.44}     & \miscval{99.32}   & \miscval{98.80}   & \miscval{99.62}    & \miscval{99.30}  \\ \midrule
            \nothinking{} \negatives{} \online{}   & (\S\ref{sec: think-toggle}) GRPO-Instruct &               & \miscval{100.00}    & \miscval{100.00}  & \miscval{100.00}  & \miscval{100.00}   & \miscval{100.00} \\
            \thinking{} \negatives{} \online{}     & (\S\ref{sec: think-toggle}) GRPO-Think-4k &               & \miscval{98.64}     & \miscval{99.17}   & \miscval{97.13}   & \miscval{99.05}    & \miscval{98.50}  \\
            \thinking{} \negatives{} \online{}     & (\S\ref{sec: think-toggle}) GRPO-Think-8k &               & \miscval{98.43}     & \miscval{98.43}   & \miscval{98.58}   & \miscval{98.61}    & \miscval{98.51}  \\
            \thinking{} \negatives{} \noonline{}   & (\S\ref{sec: online-toggle}) DPO-Think    & 7B            & \miscval{94.08}     & \miscval{95.62}   & \miscval{91.98}   & \miscval{95.19}    & \miscval{94.22}  \\
            \thinking{} \negatives{} \semionline{} & (\S\ref{sec: online-toggle}) BO-GRPO      &               & \miscval{95.56}     & \miscval{97.22}   & \miscval{92.23}   & \miscval{96.01}    & \miscval{95.26}  \\
            \thinking{} \nonegatives{} \online{}   & (\S\ref{sec: neg-toggle}) RAFT            &               & \miscval{83.53}     & \miscval{84.67}   & \miscval{85.56}   & \miscval{85.18}    & \miscval{84.74}  \\
            \thinking{} \negatives{} \online{}     & (\S\ref{sec: results})~~~GRPO-Think-16k   &               & \miscval{92.91}     & \miscval{95.06}   & \miscval{90.62}   & \miscval{94.20}    & \miscval{93.20}  \\ \midrule
            \nothinking{} \negatives{} \online{}   & (\S\ref{sec: think-toggle}) GRPO-Instruct &               & \miscval{99.85}     & \miscval{99.88}   & \miscval{99.97}   & \miscval{99.71}    & \miscval{99.85}  \\
            \thinking{} \negatives{} \online{}     & (\S\ref{sec: think-toggle}) GRPO-Think-4k &               & \miscval{95.56}     & \miscval{97.16}   & \miscval{96.30}   & \miscval{95.79}    & \miscval{96.20}  \\
            \thinking{} \negatives{} \online{}     & (\S\ref{sec: think-toggle}) GRPO-Think-8k &               & \miscval{99.88}     & \miscval{99.94}   & \miscval{100.00}  & \miscval{99.90}    & \miscval{99.93}  \\
            \thinking{} \negatives{} \noonline{}   & (\S\ref{sec: online-toggle}) DPO-Think    & 14B           & \miscval{99.54}     & \miscval{99.63}   & \miscval{99.01}   & \miscval{99.62}    & \miscval{99.45}  \\
            \thinking{} \negatives{} \semionline{} & (\S\ref{sec: online-toggle}) BO-GRPO      &               & \miscval{99.04}     & \miscval{99.44}   & \miscval{98.37}   & \miscval{99.11}    & \miscval{98.99}  \\
            \thinking{} \nonegatives{} \online{}   & (\S\ref{sec: neg-toggle}) RAFT            &               & \miscval{97.16}     & \miscval{97.62}   & \miscval{96.88}   & \miscval{96.74}    & \miscval{97.10}  \\
            \thinking{} \negatives{} \online{}     & (\S\ref{sec: results})~~~GRPO-Think-16k   &               & \miscval{97.72}     & \miscval{98.03}   & \miscval{90.44}   & \miscval{97.91}    & \miscval{96.03}  \\ \bottomrule
        \end{tabular}
    }
    \label{tab: parse-rates}
\end{table}

\subsection{Results from External Benchmarks and Downstream Integration}
\label{sec: downstream-results}
We report the detailed results from our external benchmarks and downstream integration experiments in \cref{tab: downstream_performance}. RM-Bench~\citep{liuRMBenchBenchmarkingReward2024} is a general-purpose benchmark that probes a reward model's sensitivity to subtle content differences and its robustness to stylistic biases (e.g., length or formatting) across chat, math, code, and safety domains. We report its overall score (RMBench-O) and its code-specific subset (RMBench-C). Themis-CodeRewardBench (CRB;~\citealp{DBLP:journals/corr/abs-2605-00754}) is a benchmark for coding reward models, spanning several auxiliary code-related tasks with multi-criteria scoring. We report its overall score (CRB-O). Since both benchmarks use a paired-accuracy formulation over preference pairs, they let us situate our verifiers against established reward-modeling evaluations beyond the \workname{} testbed. As discussed in \cref{sec: downstream}, we find that the performance trends observed in our controlled testbed are consistent with both, downstream integration and external benchmarks. This validates the utility of our testbed for stress-testing verifiers as best-of-N selectors in a more cost-effective manner.
\begin{table}[!t]
    \centering
    \caption{\textbf{Results on external benchmarks and downstream BoN integration.} We report the overall scores for RM-Bench and CRB, and the scores on the code split for RM-Bench. Subscripts are 95\% confidence intervals.}
    \label{tab: downstream_performance}
    \tt
    \small
    \setlength{\tabcolsep}{6pt}\renewcommand{\arraystretch}{1.3}
    \scalebox{0.8}{
        \begin{tabular}{lccc|cccc}
            \multicolumn{8}{c}{\thinking{}~\texttt{\textbf{Thinking}} \quad \negatives{}~\texttt{\textbf{Negatives}} \quad \online{}~\texttt{\textbf{Online}} \quad \semionline{}~\texttt{\textbf{Batch-online}}}                    \\
            \toprule
            \textbf{Algorithm}                                 & \textbf{Size}         & \textbf{ListAcc}     & \textbf{\ktau{}}     & \textbf{BoN}          & \textbf{CRB-O}        & \textbf{RMBench-O}    & \textbf{RMBench-C}    \\\midrule
            \nothinking{} \negatives{} \online{} GRPO-Instruct & \multirow{7}{*}{1.5B} & \scvalc{35.41}{0.71} & \rcvalc{7.82}{2.27}  & \bonvalc{16.59}{2.24} & \extvalc{56.52}{1.04} & \extvalc{62.96}{1.55} & \extvalc{50.19}{3.80} \\
            \thinking{} \negatives{} \online{} GRPO-Think-4k   &                       & \scvalc{37.76}{0.52} & \rcvalc{9.81}{2.40}  & \bonvalc{18.10}{2.32} & \extvalc{58.19}{1.02} & \extvalc{60.64}{1.65} & \extvalc{52.97}{3.63} \\
            \thinking{} \negatives{} \online{} GRPO-Think-8k   &                       & \scvalc{42.50}{0.52} & \rcvalc{15.09}{2.43} & \bonvalc{19.81}{2.40} & \extvalc{58.20}{1.02} & \extvalc{63.57}{1.72} & \extvalc{53.12}{4.08} \\
            \thinking{} \negatives{} \noonline{} DPO-Think     &                       & \scvalc{21.08}{0.32} & \rcvalc{13.48}{2.61} & \bonvalc{18.39}{2.33} & \extvalc{22.50}{0.86} & \extvalc{33.27}{1.63} & \extvalc{3.65}{1.31}  \\
            \thinking{} \negatives{} \semionline{} BO-GRPO     &                       & \scvalc{38.96}{0.46} & \rcvalc{10.97}{2.49} & \bonvalc{19.53}{2.39} & \extvalc{58.59}{1.02} & \extvalc{63.68}{1.72} & \extvalc{53.12}{4.16} \\
            \thinking{} \nonegatives{} \online{} RAFT          &                       & \scvalc{32.32}{0.39} & \rcvalc{11.96}{2.47} & \bonvalc{18.29}{2.33} & \extvalc{49.11}{1.04} & \extvalc{62.92}{1.69} & \extvalc{53.90}{4.10} \\
            \thinking{} \negatives{} \online{} GRPO-Think-16k  &                       & \scvalc{44.34}{0.55} & \rcvalc{19.11}{2.42} & \bonvalc{19.91}{2.41} & \extvalc{57.86}{1.02} & \extvalc{64.80}{1.78} & \extvalc{53.27}{4.29} \\\midrule
            \nothinking{} \negatives{} \online{} GRPO-Instruct & \multirow{7}{*}{7B}   & \scvalc{50.09}{0.75} & \rcvalc{25.90}{2.23} & \bonvalc{18.77}{2.36} & \extvalc{66.79}{0.98} & \extvalc{70.96}{1.61} & \extvalc{57.31}{4.65} \\
            \thinking{} \negatives{} \online{} GRPO-Think-4k   &                       & \scvalc{51.31}{0.61} & \rcvalc{26.55}{2.37} & \bonvalc{20.95}{2.46} & \extvalc{66.17}{0.98} & \extvalc{70.50}{1.59} & \extvalc{55.70}{4.25} \\
            \thinking{} \negatives{} \online{} GRPO-Think-8k   &                       & \scvalc{56.79}{0.51} & \rcvalc{28.40}{2.31} & \bonvalc{22.27}{2.51} & \extvalc{68.54}{0.96} & \extvalc{76.65}{1.61} & \extvalc{61.55}{4.78} \\
            \thinking{} \negatives{} \noonline{} DPO-Think     &                       & \scvalc{55.86}{0.50} & \rcvalc{28.71}{2.30} & \bonvalc{22.75}{2.52} & \extvalc{67.91}{0.98} & \extvalc{77.63}{1.59} & \extvalc{63.99}{4.74} \\
            \thinking{} \negatives{} \semionline{} BO-GRPO     &                       & \scvalc{55.47}{0.54} & \rcvalc{28.43}{2.30} & \bonvalc{22.09}{2.50} & \extvalc{69.38}{0.96} & \extvalc{76.27}{1.61} & \extvalc{60.23}{4.66} \\
            \thinking{} \nonegatives{} \online{} RAFT          &                       & \scvalc{52.69}{0.52} & \rcvalc{29.35}{2.33} & \bonvalc{21.99}{2.49} & \extvalc{64.21}{1.00} & \extvalc{77.51}{1.59} & \extvalc{64.13}{4.70} \\
            \thinking{} \negatives{} \online{} GRPO-Think-16k  &                       & \scvalc{65.03}{0.53} & \rcvalc{41.07}{2.08} & \bonvalc{22.84}{2.53} & \extvalc{70.25}{0.96} & \extvalc{79.85}{1.59} & \extvalc{71.05}{4.63} \\\midrule
            \nothinking{} \negatives{} \online{} GRPO-Instruct & \multirow{7}{*}{14B}  & \scvalc{54.25}{0.71} & \rcvalc{29.30}{2.17} & \bonvalc{18.58}{2.34} & \extvalc{72.34}{0.94} & \extvalc{74.43}{1.65} & \extvalc{62.28}{4.98} \\
            \thinking{} \negatives{} \online{} GRPO-Think-4k   &                       & \scvalc{62.72}{0.61} & \rcvalc{34.67}{2.13} & \bonvalc{22.27}{2.51} & \extvalc{81.46}{0.80} & \extvalc{83.39}{1.45} & \extvalc{67.54}{4.92} \\
            \thinking{} \negatives{} \online{} GRPO-Think-8k   &                       & \scvalc{68.95}{0.57} & \rcvalc{39.70}{2.06} & \bonvalc{22.94}{2.53} & \extvalc{83.90}{0.76} & \extvalc{83.80}{1.45} & \extvalc{70.18}{4.84} \\
            \thinking{} \negatives{} \noonline{} DPO-Think     &                       & \scvalc{73.90}{0.51} & \rcvalc{41.49}{2.00} & \bonvalc{23.79}{2.57} & \extvalc{85.79}{0.73} & \extvalc{85.23}{1.41} & \extvalc{71.10}{4.98} \\
            \thinking{} \negatives{} \semionline{} BO-GRPO     &                       & \scvalc{75.24}{0.50} & \rcvalc{42.19}{1.98} & \bonvalc{23.79}{2.57} & \extvalc{85.69}{0.73} & \extvalc{86.97}{1.33} & \extvalc{74.12}{4.78} \\
            \thinking{} \nonegatives{} \online{} RAFT          &                       & \scvalc{67.23}{0.52} & \rcvalc{41.24}{2.07} & \bonvalc{23.32}{2.54} & \extvalc{85.70}{0.73} & \extvalc{84.30}{1.43} & \extvalc{69.15}{4.84} \\
            \thinking{} \negatives{} \online{} GRPO-Think-16k  &                       & \scvalc{80.50}{0.46} & \rcvalc{52.03}{1.82} & \bonvalc{24.83}{2.60} & \extvalc{85.27}{0.74} & \extvalc{88.61}{1.27} & \extvalc{80.80}{4.49} \\\bottomrule
        \end{tabular}
    }
\end{table}

\subsection{Transferrability to Other Model Families}
\label{sec: olmo-results}
Throughout the experiments in the main text, we conducted experiments on the \texttt{DeepSeek-R1-Distill-Qwen2.5 (DSQwen)} model family. The reason behind this choice was because they were distilled to generate intermediate thinking traces from \texttt{DeepSeek-R1}, allowing a neat comparison to the corresponding \texttt{Qwen2.5} model. This is especially valuable because recent LLMs are trained to be hybrid thinking/instruct models whose reasoning effort can be difficult to control~\citep{yang2025qwen3technicalreport}. The validity of our results beyond this choice is a natural question, which we aim to address in this section. Specifically, we train the \texttt{Olmo3-7B} models~\citep{olmo2026olmo3}, which similarly offer distinct \textit{-Think} and \textit{-Instruct} models. We reproduce the three main comparisons in \cref{sec: results} in \cref{tab: olmo-vs-dsq}.

\begin{table*}[!t]
    \centering
    \caption{\textbf{Results for \texttt{Olmo3-7B}.} We report \texttt{ListAcc} and \ktau{} scores for BoN and RL, respectively, across the three ablation axes. \texttt{Olmo3} performs markedly better than \texttt{DeepSeek-R1-Distill-Qwen2.5} on most evaluations. Thinking traces remain the most impactful axis in both families at \texttt{7B}. Subscripts are 95\% confidence intervals.}
    \small
    \tt
    \setlength{\tabcolsep}{3pt}
    \scalebox{0.85}{
        \begin{tabular}{ll|w{c}{1.3cm}w{c}{1.3cm}|w{c}{1.3cm}w{c}{1.3cm}|w{c}{1.3cm}w{c}{1.3cm}|w{c}{1.3cm}w{c}{1.3cm}|w{c}{1.1cm}w{c}{1.1cm}}
            \multicolumn{12}{c}{\thinking{}~\texttt{\textbf{Thinking}} \quad \negatives{}~\texttt{\textbf{Negatives}} \quad \online{}~\texttt{\textbf{Online}}}                                                                                                                                                                                                                                                        \\
            \toprule
                                                   & \multirow{2}{*}{\textbf{Method}} & \multicolumn{2}{c|}{\textbf{\heldout{}}} & \multicolumn{2}{c|}{\textbf{\rqone{}}} & \multicolumn{2}{c|}{\textbf{\rqtwo{}}} & \multicolumn{2}{c|}{\textbf{\rqfour{}}} & \multicolumn{2}{c}{\textbf{Average}}                                                                                                                    \\
            \cmidrule(lr){3-4} \cmidrule(lr){5-6} \cmidrule(lr){7-8} \cmidrule(lr){9-10} \cmidrule(lr){11-12}
                                                   &                                  & \texttt{BoN}                             & \texttt{RL}                            & \texttt{BoN}                           & \texttt{RL}                             & \texttt{BoN}                         & \texttt{RL}          & \texttt{BoN}         & \texttt{RL}          & \texttt{BoN}         & \texttt{RL}          \\ \midrule
            \nothinking{}\,\negatives{}\,\online{} & GRPO-Instruct                    & \scvalc{66.03}{0.71}                     & \rcvalc{36.45}{2.03}                   & \scvalc{58.53}{0.68}                   & \rcvalc{28.13}{2.21}                    & \scvalc{45.13}{0.82}                 & \rcvalc{11.35}{2.34} & \scvalc{53.96}{0.37} & \rcvalc{27.16}{2.18} & \scvalc{57.26}{0.64} & \rcvalc{25.77}{2.19} \\
            \thinking{}\,\nonegatives{}\,\online{} & RAFT                             & \scvalc{77.29}{0.55}                     & \rcvalc{47.78}{1.88}                   & \scvalc{68.68}{0.59}                   & \rcvalc{46.09}{2.04}                    & \scvalc{65.50}{0.63}                 & \rcvalc{24.46}{2.31} & \scvalc{71.32}{0.28} & \rcvalc{43.33}{2.09} & \scvalc{70.69}{0.51} & \rcvalc{40.42}{2.08} \\
            \thinking{}\,\negatives{}\,\noonline{} & DPO-Think                        & \scvalc{75.60}{0.57}                     & \rcvalc{46.86}{1.95}                   & \scvalc{67.07}{0.60}                   & \rcvalc{44.97}{2.12}                    & \scvalc{64.93}{0.66}                 & \rcvalc{23.99}{2.36} & \scvalc{68.31}{0.27} & \rcvalc{41.22}{2.14} & \scvalc{68.97}{0.52} & \rcvalc{39.26}{2.15} \\
            \thinking{}\,\negatives{}\,\online{}   & GRPO-Think-16k                   & \scvalc{86.23}{0.47}                     & \rcvalc{55.41}{1.71}                   & \scvalc{74.12}{0.52}                   & \rcvalc{52.00}{1.99}                    & \scvalc{67.74}{0.69}                 & \rcvalc{25.53}{2.28} & \scvalc{76.23}{0.24} & \rcvalc{48.74}{1.79} & \scvalc{76.08}{0.48} & \rcvalc{45.42}{1.94} \\ \bottomrule
        \end{tabular}
    }
    \label{tab: olmo-vs-dsq}
\end{table*}

\texttt{Olmo3} performs markedly better than \texttt{DSQwen} across all three axes, with the \texttt{7B} model being competitive with the \texttt{DSQwen-14B} and clearly ahead of the \texttt{DSQwen-7B}. \texttt{GRPO-Think} recipe remains the best verifier on every slice and both metrics. Similar to the main text, thinking is the most impactful axis, with \texttt{GRPO-Instruct} being the worst-performing ablation for both model families. The gap to GRPO-Think is wider for the stronger \texttt{Olmo3} models, consistent with our findings in \cref{sec: think-toggle}. Unlike \texttt{DSQwen}, \texttt{RAFT} narrowly outperforms \texttt{DPO-Think} for \texttt{Olmo3}, but the difference between the two methods is small for both model families. Thus, the results from \texttt{Olmo3-7B} suggest that our findings are likely not limited to our choice of using \texttt{DeepSeek-R1-Distill-Qwen2.5} models in the main text.

\newpage
\section{Prompt Templates}
\label{sec: prompt_templates}
\begin{prompt}{Default training prompt}
    \small \ttfamily
    You are an expert judge of coding problems. Given a coding problem and multiple candidate solutions, your task is to evaluate the correctness of each solution based on the problem description. Your evaluation should be based solely on the functional correctness of the code. It is guaranteed that exactly one of the candidates is completely correct. Here is the coding question followed by the candidate solutions:

    [QUESTION]
    \textcolor{promptblue}{\{question\}}
    [/QUESTION]
    \newline\newline

    [CANDIDATE\_A]
    \textcolor{promptblue}{\{code\_A\}}
    [/CANDIDATE\_A]
    \newline\newline
    [CANDIDATE\_B]
    ...

    Indicate your choice of candidate only by responding with one of the following options: \textcolor{promptblue}{\{valid\_options\}}. Enclose your final answer in the format $\backslash$boxed\{X\}, where X is your chosen option among the candidates. Do not provide any additional text. Your response should be exactly one of the options enclosed within $\backslash$boxed\{\}, without any extra characters or spaces. Anything else will be considered invalid.
\end{prompt}

\begin{prompt}{\texttt{GRPO-Instruct} training prompt}
    \small \ttfamily
    You are an expert judge of coding problems. Given a coding problem and multiple candidate solutions, your task is to evaluate the correctness of each solution based on the problem description.
    Your evaluation should be based solely on the functional correctness of the code. It is guaranteed that exactly one of the candidates is completely correct.
    Indicate your choice of candidate by responding with one of the following options: \textcolor{promptblue}{\{valid\_options\}}. Your response should be in the following format:
    \newline
    Analysis: <Your step-by-step reasoning here>
    \newline
    Final Answer: $\backslash$boxed\{X\}, where X is your chosen option among the candidates.
    \newline
    \newline
    Here is the coding question followed by the candidate solutions:
    \newline
    [QUESTION]
    \textcolor{promptblue}{\{question\}}
    [/QUESTION]
    \newline\newline
    [CANDIDATE\_A]
    \textcolor{promptblue}{\{code\_A\}}
    [/CANDIDATE\_A]
    \newline\newline
    [CANDIDATE\_B]
    ...
    \newline
    Your response should be exactly in the specified format, without any extra characters or spaces.
    Anything else will be considered invalid.
    \newline
\end{prompt}

\begin{prompt}{ListSc and PairSc training prompt}
    \small \ttfamily
    You are an expert judge of coding problems. Given a coding problem and two candidate solutions, your task is to evaluate the correctness of each solution based on the problem description. Your evaluation should be based solely on the functional correctness of the code. It is guaranteed that exactly one of the candidates is completely correct. Here is the coding question followed by the candidate solutions:
    \newline
    [QUESTION]
    \textcolor{promptblue}{\{question\}}
    [/QUESTION]
    \newline\newline
    [CANDIDATE\_A]
    \textcolor{promptblue}{\{code\_A\}}
    [/CANDIDATE\_A]
    \newline\newline
    [CANDIDATE\_B]
    ...
    \newline
    Assign a score between 0 and 10 to EACH candidate, with 10 indicating a perfect solution that passes all test cases, 5 indicating a solution that would pass some test cases but not all, and 0 indicating a solution that fails all test cases. Output your final answer in the format $\backslash$boxed\{[<score\_candidate\_A>,<score\_candidate\_B>, <score\_candidate\_C>, ...]\} for each input candidate. Do not provide any additional text. Your response should be a list of numbers between 0 and 10, enclosed within $\backslash$boxed\{\}, without any extra characters or spaces. Anything else will be considered invalid.
\end{prompt}

\begin{prompt}{Python code generation prompt}
    \small \ttfamily
    You are an expert Python programmer. You will be given a question (problem specification) and will generate a correct Python program that matches the specification and passes all tests. Read the inputs from STDIN, solve the problem, and write the answer to STDOUT (do not directly test on the sample inputs). Enclose your code within a Python markdown block. Ensure that when the Python program runs, it reads the inputs, runs the algorithm, and writes output to STDOUT.
\end{prompt}

\begin{prompt}{C++ code generation prompt}
    \small \ttfamily
    You are an expert C++ programmer. You will be given a question (problem specification) and will generate a correct C++ program with a main function that matches the specification and passes all tests. Read the inputs from STDIN, solve the problem, and write the answer to STDOUT (do not directly test on the sample inputs). Enclose your code within a C++ markdown block. Ensure that when the C++ program runs, it reads the inputs, runs the algorithm, and writes output to STDOUT.
\end{prompt}

\begin{prompt}{Java code generation prompt}
    \small \ttfamily
    You are an expert Java programmer. You will be given a question (problem specification) and will generate a correct Java program with a public class named Main that matches the specification and passes all tests. Your class should include a public static void main(String[] args) method. Read the inputs from System.in, solve the problem, and write the answer to System.out (do not directly test on the sample inputs). Enclose your code within a Java markdown block. Ensure that when the Java program runs, it reads the inputs, runs the algorithm, and writes output to System.out.
\end{prompt}


\end{document}